\documentclass[9pt,shortpaper]{ieeetran}
\usepackage{cite}
\usepackage{amsmath,amssymb,amsfonts}
\usepackage{textcomp}
\usepackage[pdftex]{graphicx}

\usepackage{amsthm}
\usepackage{mathtools}	%
\usepackage{grffile}	%
\usepackage[tight,footnotesize]{subfigure}
\usepackage{microtype} %
\usepackage[misc]{ifsym}
\usepackage{color}
\let\labelindent\relax
\usepackage{enumitem}

\usepackage{adjustbox}
\usepackage{tikz}
\usetikzlibrary{calc,intersections,arrows.meta,bending,patterns,angles,matrix}
\usepackage{pgfplots}
\usepgfplotslibrary{fillbetween}

\usepackage{hyperref}
\hypersetup{
	colorlinks=true,
	linkcolor=blue
}

\theoremstyle{remark}
\newtheorem{remarkx}{Remark}
\newenvironment{remark}
{\pushQED{\qed}\remarkx}
{\popQED\endremarkx}
\newtheorem{examplex}{Example}
\newenvironment{example}
{\pushQED{\qed}\examplex}
{\popQED\endexamplex}
\newtheorem{counterexamplex}{Counterexample}
\newenvironment{counterexample}
{\pushQED{\qed}\counterexamplex}
{\popQED\endcounterexamplex}

\theoremstyle{definition}
\newtheorem{defn}{Definition}
\newtheorem{assump}{Assumption}

\theoremstyle{plain}
\newtheorem{theorem}{Theorem}
\newtheorem{lemma}{Lemma}
\newtheorem{coroll}{Corollary}
\newtheorem{prop}{Proposition}

\newtheorem*{claim*}{Claim}

\newcommand{\defeq}{:=} %
\newcommand{\diag}[1]{{\rm diag}\{ #1\}}
\newcommand{\dt}{\frac{{\rm d}}{{\rm d}t}}
\newcommand{\dist}{{\rm dist}}
\newcommand{\matr}[1]{\begin{bmatrix} #1 \end{bmatrix}}
\newcommand{\transpose}[1]{#1^\top}
\newcommand{\inv}[1]{#1^{-1}}	%
\newcommand{\norm}[1]{\left\lVert#1\right\rVert}
\newcommand{\chiup}{\raisebox{2pt}{$\chi$}}
\newcommand{\vf}{\chiup}
\newcommand{\identity}{{\rm id}}
\newcommand{\set}[1]{\mathcal{#1}}
\newcommand{\mbr}[1][{}]{\mathbb{R}^{#1}}	%
\newcommand{\manifold}{\set{M}}
\newcommand{\gradient}{\,\mathrm{grad}\,}
\newcommand{\orthoterm}{\bot_{e}}
\newcommand{\difrntl}{d\,}	%

\newcommand{\scalemath}[2]{\scalebox{#1}{\mbox{\ensuremath{\displaystyle #2}}}}

\begin{document}
\title{The Domain of Attraction of the Desired Path \\ in Vector-field Guided Path Following}
\author{Weijia Yao, \IEEEmembership{Member, IEEE}, Bohuan Lin, Brian D. O. Anderson, \IEEEmembership{Life fellow, IEEE}, and Ming Cao, \IEEEmembership{Fellow, IEEE}
	\thanks{}
}

\maketitle

\begin{abstract}
	In the vector-field guided path-following problem, a sufficiently smooth vector field is designed such that its integral curves converge to and move along a one-dimensional geometric desired path. The existence of singular points where the vector field vanishes creates a topological obstruction to global convergence to the desired path and some associated topological analysis has been conducted in \cite{yao2021topo}. In this paper, we strengthen the result in \cite[Theorem 2]{yao2021topo} by showing that the domain of attraction of the desired path, which is a compact asymptotically stable one-dimensional embedded submanifold of an $n$-dimensional ambient manifold $\manifold$, is homeomorphic to $\mbr[n-1] \times \mathbb{S}^1$, and not just homotopy equivalent to $\mathbb{S}^1$ as shown in \cite[Theorem 2]{yao2021topo}. This result is extended for a $k$-dimensional compact manifold for $k \ge 2$.
\end{abstract}

\begin{IEEEkeywords}
	Path following, domain of attraction, nonlinear systems
\end{IEEEkeywords}

\section{Introduction}
In mobile robotics and some control applications, it is fundamental for trajectories of a dynamical system  to converge to and accurately follow a predefined one-dimensional desired path. In this path-following problem, the desired path is usually given as a geometric or set-theoretic object rather than a temporal function or the output of an exosystem (c.f. trajectory tracking or output regulation problems). Among different path-following algorithms, those based on (guiding) vector fields have been verified to be superior in their performance in benchmark tests to follow a circle or a straight line \cite{Sujit2014}. In these vector-field guided path-following algorithms, a vector field is designed such that its integral curves converge to and continue to move along the desired path as $t \to \infty$.

Despite the advantages of vector-field guided path-following algorithms, a major problematic issue is the existence of singular points where the vector field vanishes. This issue is the motivation for a topological analysis of these algorithms. In our previous work \cite{yao2021topo}, under the common assumption that the desired path is homeomorphic to the unit circle $\mathbb{S}^1$, a number of examples are included demonstrating the existence of a singular set and initial conditions from which convergence occurs to the singular set, even though it is non-attractive. In all these examples, the set of initial conditions from which convergence occurs to the desired path (i.e., the domain of attraction) is homotopy equivalent to $\mathbb{S}^1$ \cite[Theorem 2]{yao2021topo}, and in retrospect, is seen to also be homeomorphic to $\mathbb{R}^{n-1}\times \mathbb{S}^1$, where $n$ is the dimension of the ambient manifold $\manifold$. But since there are topological objects which are homotopy equivalent to $\mathbb{S}^1$, but not homeomorphic to $\mathbb{R}^{n-1}\times \mathbb{S}^1$, it is plain that the earlier result may well be inadequate; indeed, we shall show that the homeomorphic property necessarily holds, using tools more powerful than those of \cite{yao2021topo}.

The aforementioned desired path is a \emph{compact asymptotically stable embedded} submanifold, denoted by $\set{P}$ for convenience, in an ambient finite-dimensional manifold. Some studies also characterize the domain of attraction of $\set{P}$. In \cite[Chapter V, Lemma 3.2]{bhatia2002stability}, it has been shown that the intersection of an $\epsilon$-neighborhood of $\set{P}$ and some sublevel set of a Lyapunov function is a deformation retract \cite[p. 200]{lee2010topologicalmanifolds} of the domain of attraction of $\set{P}$. This result has been strengthened in \cite{moulay2010topological,bernuau2019topological}, which prove that $\set{P}$ itself is actually a deformation retract of its domain of attraction, and thus $\set{P}$ and its domain of attraction are homotopy equivalent. In \cite[Theorem 3.4]{wilson1967structure}, it is claimed that $\set{P}$ is diffeomorphic to a tubular neighborhood of itself. %

The importance of topological analysis of domains of attraction is also revealed in motion planning problems. For motion-planning algorithms using the negative gradient of a potential/navigation function, \cite{koditschek1990robot} demonstrated that the shape of the configuration space of the problem could affect the ability to plan a path from a starting point to a destination point; typically there are particular initial points in the space from which the motion planning task cannot be achieved without major modification (e.g. introduction of discontinuity). In this paper and its earlier companion \cite{yao2021topo}, we show that such problems could be significantly worse, in the sense that the sets of such points could have positive measure rather than measure zero, or even being just finite sets. This is a major driver for the seeking of a general understanding (using largely topological concepts) of the situation in which such difficulties can arise. 

In this paper, we execute a major strengthening of the result in \cite[Theorem 2]{yao2021topo}. We show that the domain of attraction of the desired path $\set{P}$ is not just homotopy equivalent to $\mathbb{S}^1$ (as shown in \cite[Theorem 2]{yao2021topo}), but is actually homeomorphic to $\mbr[n-1] \times \mathbb{S}^1$. It is well known that there is no systematic way to find the domain of attraction. Therefore, the derivation of this result is highly challenging, and draws on a number of ideas of differential topology for its detailed proof (which, for the sake of the reader's convenience, is confined as far as possible to the appendix). Note that the result in \cite[Theorem 2]{yao2021topo} relies on \cite{moulay2010topological}, of which the proof involves the properties of Lyapunov functions, the continuity of the first hitting time and the construction of homotopies. However, in this paper, except for the Lyapunov functions, we use distinctively different and necessarily more powerful proof techniques, including the local triviality near a compact regular level set (see Proposition \ref{note_prop_new} and Lemma \ref{lemma_trivialnormalbundle}), the study of properties of a vector field on another topological space (see Lemma \ref{note_lemma3}), the use of the exit set $\set{W}^{e}$ and related sets (see Lemmas \ref{note_lemma3} and \ref{note_lemma5}) and the construction of quotient maps (see Proposition \ref{note_prop3}).

\section{Problem formulation} \label{sec_problemform}
In this section, we provide preliminaries on the theory of topology and manifolds, and review the guiding vector field defined on an $n$-dimensional Riemannian manifold $\manifold$ with the Riemannian metric denoted by $g$ in \cite{yao2021topo}, after which we formulate the problem.  

\subsection{Preliminaries} \label{subsec_preliminary}

Throughout the paper, we use calligraphic capital letters for sets (e.g., $\set{A}$) and regular lowercase letters or Greek letters (e.g., $f$, $\Gamma$) for mappings. ``w.r.t.'' is the abbreviation for ``with respect to''.

Suppose $\set{X}, \set{Y}$ are two topological spaces. A \emph{component of $\set{X}$} is a maximal nonempty connected subset of $\set{X}$ (i.e., a nonempty connected subset that is not properly contained in any other connected subset of $\set{X}$). A subset of $\set{X}$ is \emph{precompact in $\set{X}$} if its closure in $\set{X}$ is compact.	A \emph{homeomorphism} (\emph{diffeomorphism} resp.) is a continuous (smooth resp.) bijection $f: \set{X} \to \set{Y}$ which has a continuous (smooth resp.) inverse. If there exists a homeomorphism between $\set{X}$ and $\set{Y}$, then $\set{X}$ and $\set{Y}$ are called \emph{homeomorphic}, denoted by $\set{X} \approx \set{Y}$. A \emph{homotopy equivalence} between $\set{X}$ and $\set{Y}$ is a pair of continuous maps $g: \set{X} \to \set{Y}$ and $h: \set{Y} \to \set{X}$ such that $g \circ h$ is homotopic to the identity map $\identity_{\set{Y}}: \set{Y} \to \set{Y}$, and $h \circ g$ is homotopic to the identity map $\identity_{\set{X}}: \set{X} \to \set{X}$. If this pair exists, then $\set{X}$ and $\set{Y}$ are \emph{homotopy equivalent}. It is obvious that a homeomorphism is a special case of a homotopy equivalence obtained by letting $g = h^{-1}$, and this also leads to the stronger results that $g \circ h$ (and $h \circ g$  resp.) is equal to (rather than homotopic to) $\identity_{\set{Y}}$ (and $\identity_{\set{X}}$ resp.). With $\set{A} \subseteq \set{X}$, a continuous map $r: \set{X} \to \set{A}$ is a \emph{retraction} if the restriction of $r$ to $\set{A}$ is the identity map of $\set{A}$, or equivalently if $r \circ \iota_{\set{A}} = \identity_{\set{A}}$, where $\iota_{\set{A}}: \set{A} \to \set{X}$ is the inclusion map. Given functions $f: \set{X} \to \set{X}'$ and $g: \set{Y} \to \set{Y}'$, the map $f \times g: \set{X} \times \set{Y} \to \set{X}' \times \set{Y}'$ is defined by $(f \times g)(x,y)=(f(x), g(y))$ for $x \in \set{X}$ and $y \in \set{Y}$. 

Let $\set{M}$ and $\set{N}$ be smooth manifolds. The \emph{tangent space} $T_{\xi} \set{M}$ to $\set{M}$ at ${\xi} \in \set{M}$ is a vector space consisting of maps (called \emph{derivations}) $f_{\xi}: C^{\infty}({\xi}) \to \mbr[]$ which satisfy the linearity and product rules, where $C^{\infty}({\xi})$ is the set of smooth real-valued functions defined on an open neighborhood of ${\xi}$. Given a smooth map $f: \set{M} \to \set{N}$, the \emph{tangent (or differential) map of $f$ at ${\xi} \in \set{M}$} is denoted by $\difrntl f_{{\xi}}: T_{\xi} \set{M} \to T_{f({\xi})} \set{N}$. By definition, there holds that $(\difrntl f_{{\xi}} (X_{\xi})) (g) = X_{\xi} (g \circ f)$ for $X_{\xi} \in T_{\xi} \set{M}$ and $g \in C^\infty(f({\xi}))$. If the subscript ${\xi}$ is omitted, then $\difrntl f: T \set{M} \to T \set{N}$ is a map defined at any point ${\xi} \in \set{M}$, where $T \set{M}$ is the \emph{tangent bundle} $T \set{M} \defeq \sqcup_{{\xi} \in \set{M}} T_{\xi} \set{M}$ with $\sqcup$ being the disjoint union, and $T \set{N}$ is defined similarly. The map $f$ is a \emph{submersion on $\set{M}$} if for any ${\xi} \in \set{M}$, the tangent map $\difrntl f_{{\xi}}$ at ${\xi}$ is surjective. If the tangent map $\difrntl f_{{\xi}}$ at ${\xi}$ is surjective, then ${\xi}$ is a \emph{regular point}. If for every ${\xi} \in \inv{f}(q)$,  the tangent map $\difrntl f_{{\xi}}$ is surjective, then $q \in \set{N}$ is a \emph{regular value of $f$}. Refer to \cite{lee2015introduction} for more detail about differential manifolds.

The distance between a point ${\xi} \in \manifold$ and a submanifold $\set{N} \subseteq \manifold$ is defined as $\dist({\xi}, \set{N}) = \dist(\set{N}, {\xi}) \defeq \inf \{ d({\xi}, q) : q \in \set{N} \} $, where $d(\cdot, \cdot)$ is the Riemannian distance between two points in $\manifold$. The distance between two submanifolds $\set{N}, \set{N}' \subseteq \manifold$ is defined as $\dist(\set{N}, \set{N}') = \dist(\set{N}', \set{N}) \defeq \inf \{d({\xi},q) : {\xi} \in \set{N}, q \in \set{N}' \}$. The norm $\norm{v}$ of a tangent vector $v \in T_{\xi} \manifold$ is defined by $ \norm{v} \defeq \langle v, v \rangle_g ^{1/2}$, where $\langle \cdot, \cdot \rangle_g$ is the inner product of tangent vectors in $T_{\xi} \manifold$ w.r.t. the Riemannian metric $g$.

\subsection{Vector-field guided path following on Riemannian manifolds}
The desired path $\set{P} \subseteq \manifold$ is described as the intersection of zero-level sets of twice continuously differentiable functions ${e}_i: \manifold \to \mbr[]$, $i=1,\dots,n-1$. Namely,
$
	\set{P} \defeq \{\xi \in \manifold : {e}_i(\xi)=0, i=1,\dots,n-1 \},
$
where the functions ${e}_i$ are termed \emph{surface functions} for convenience. This description of the desired path $\set{P}$ enables one to define the \emph{path-following error} $e: \manifold \to \mbr[n-1]$ by stacking all ${e}_i$ functions together; that is,
\begin{equation} \label{eqe}
	e(\xi) = \transpose{ ({e}_1(\xi) , \cdots , {e}_{n-1}(\xi))},
\end{equation}
The reason for calling it path-following error is that, using this notion, the desired path $\set{P}$ is equivalent to 
$
	\set{P} = \{ \xi \in \manifold : e(\xi)=0 \}.
$
Using the surface functions ${e}_i$, the guiding vector field $\vf: \manifold \to T \manifold$ defined on the manifold $\manifold$ is
\begin{equation} \label{new_gvf}
	\vf(\xi) = \orthoterm(\xi) - \sum_{i=1}^{n-1} k_i {e}_i(\xi) \gradient {e}_i(\xi), \tag{GVF-M}
\end{equation}
for any point $\xi \in \manifold$, where $k_i$ are positive constants, $\gradient {e}_i(\cdot) \in T_{(\cdot)} \manifold$ are the gradient vectors of the surface functions ${e}_i$ on the manifold $\manifold$, and $\orthoterm(\cdot) \in T_{(\cdot)} \manifold$ is a generalization to the manifold $\manifold$ of the wedge product of all the gradient vectors $\gradient {e}_i(\cdot)$ \cite[Proposition 2]{yao2021topo}. The term $\orthoterm(\xi)$ is orthogonal to each of the gradient $\gradient {e}_i(\xi)$, the same as its counterpart defined in the Euclidean space, as formally stated below.
\begin{lemma}[\hspace{1sp}{\cite[Lemma 1]{yao2021topo}}] \label{lemma1} 
	There holds that 	
	$
	\langle \orthoterm(\xi), \gradient {e}_i(\xi) \rangle_g = 0
	$
	for $i=1,\dots, n-1$, and $\xi \in \manifold$, 
\end{lemma} 
The first and second terms on the right of \eqref{new_gvf} induce motion along the path, and motion towards the path, respectively. A point $\xi \in \manifold$ where the vector field vanishes (i.e., $\vf(\xi)=0$) is called a \emph{singular point}, and the set of singular points is called the \emph{singular set}, which is defined by
$
\set{C} = \{ \xi \in \manifold : \vf(\xi)=0 \}.
$

\subsection{Assumptions and problem statement}
For certain manifolds $\manifold$ and under the global dichotomy convergence condition\footnote{The condition means that all trajectories converge to either the desired path or the singular set asymptotically \cite[Section IV]{yao2021topo}.}, the existence of a singular point (or possibly a set of singular points) is unavoidable \cite[Theorem 3]{yao2021topo}. The following standing assumption is required. In effect, it is an assumption on the functions ${e}_i$. 
\begin{assump} \label{assump1}
	There are no singular points on the desired path. More precisely, $\set{C}$ is empty or otherwise there holds $\dist(\set{C}, \set{P}) > 0$. 
\end{assump}
Assumption \ref{assump1} ensures the ``regularity'' of the desired path $\set{P}$.
\begin{lemma}[\hspace{1sp}\cite{yao2021topo}, Regularity of $\set{P}$] \label{lemmanifold}
	The zero vector $0 \in \mbr[n-1]$ is a regular value of the map $e$ in \eqref{eqe}, and hence the desired path $\set{P}$ is a $C^2$ embedded submanifold in $\manifold$. 
\end{lemma}	

Given the guiding vector field $\vf: \manifold \to T \manifold$ in \eqref{new_gvf}, one investigates the solutions to the following autonomous ordinary differential equation:
\begin{equation} \label{eq1}
	\dot{\xi}(t) = \vf(\xi(t)).
\end{equation}

To obtain the topological results later, we impose the following standing assumption:
\begin{assump} \label{assump3}
	The desired path $\set{P}$ is homeomorphic to the unit circle $\mathbb{S}^1$ (i.e., $\set{P} \approx \mathbb{S}^1$). 
\end{assump}
It is shown in \cite[Theorem 2]{yao2021topo} that the domain of attraction of the desired path $\set{P}$ has the same homotopy type as the unit circle $\mathbb{S}^1$. Despite its theoretical interest, \cite[Theorem 2]{yao2021topo} gives a very rough impression of how the domain of attraction may look, since two homotopy equivalent objects may be geometrically very different; for example, $\mbr[2] \setminus \{0\}$ and $\mathbb{S}^1$ are homotopy equivalent. Another example is $\{0\}$ and $\mbr[n]$, for any positive number $n$. Suppose $\manifold=\mbr[2]$, then it is unclear, for instance, if the domain of attraction of the desired path $\set{P}$ is $\mbr[2] \setminus \{0\}$ (or $\mbr[2]$ excluding any other point). Therefore, having a stronger notion to capture the ``shape'' of the domain of attraction is highly desirable, and the problem to be solved in this paper is to obtain such a stronger notion to replace the homotopy equivalent relation between the desired path and its domain of attraction.

Observe that neither $\mbr[2] \setminus \{0\}$ and $\mathbb{S}^1$, nor $\{0\}$ and $\mbr[n]$, are \emph{homeomorphic}. Thus a candidate notion to consider is the homeomorphic relation between two topological objects. In the subsequent sections, we will strengthen \cite[Theorem 2]{yao2021topo}  by showing that the domain of attraction is homeomorphic to $\mbr[n-1] \times \mathbb{S}^1$, where $n$ is the dimension of the manifold $\manifold$ (i.e., Theorem \ref{note_thm2}).

\section{Further characterization of the domain of attraction} \label{sec_doa}
In this section, the main result, Theorem \ref{note_thm2}, about the domain of attraction of the desired path is developed. Fig. \ref{fig:overallproof} shows the overview and structure of the intermediate results which lead to Theorem \ref{note_thm2}. Overall, there are three major stages. In the first stage (consisting of Proposition \ref{note_prop_new} and Corollary \ref{coroll1}), we present the local triviality property of a neighborhood of the desired path $\set{P}$. This property reveals that the level sets in a neighborhood of the compact desired path $\set{P}$ are topologically ``similar to'' $\set{P}$, and this neighborhood turns out to be homeomorphic to a solid torus $\set{U} \times \mathbb{S}^1$, where $\set{U} \subseteq \mathbb{R}^{n-1}$. This corollary is fundamental, as the subsequent results are all based on it. In the second stage (consisting of Lemmas \ref{note_lemma3}, \ref{note_lemma4}, and \ref{note_lemma5}), we derive the analytic expressions of the so-called \emph{pre-Wazewski set} $\set{W}$, the \emph{pre-exit set} $\set{W}^{\circ}$ and the \emph{exit set} $\set{W}^{e}$ (see Definition \ref{note_def3}). These sets' physical meanings, along with some topological results, facilitate the characterization of the domain of attraction. In particular, a major technique used in Lemma \ref{note_lemma3} is defining a new vector field $\tilde{\vf}$ on the solid torus $\set{U} \times \mathbb{S}^1$, which is homeomorphic to a neighborhood of $\set{P}$ by Corollary \ref{coroll1}, and one can obtain the properties of the original vector field  $\vf$ on $\manifold$ by investigating $\tilde{\vf}$ instead. This enables one to find invariant sets and homeomorphic objects. The last stage (consisting of Proposition \ref{note_prop3} and Theorem \ref{note_thm2})  characterizes the domain of attraction by utilizing the previous topological results, the physical interpretation of $\set{W}^{e}$ and $\set{W}^{\circ}$ and the construction of quotient maps.

\subsection{Local triviality near a compact regular level set} \label{local_triviality}
\begin{figure}
	\centering
	\includegraphics[width=\linewidth]{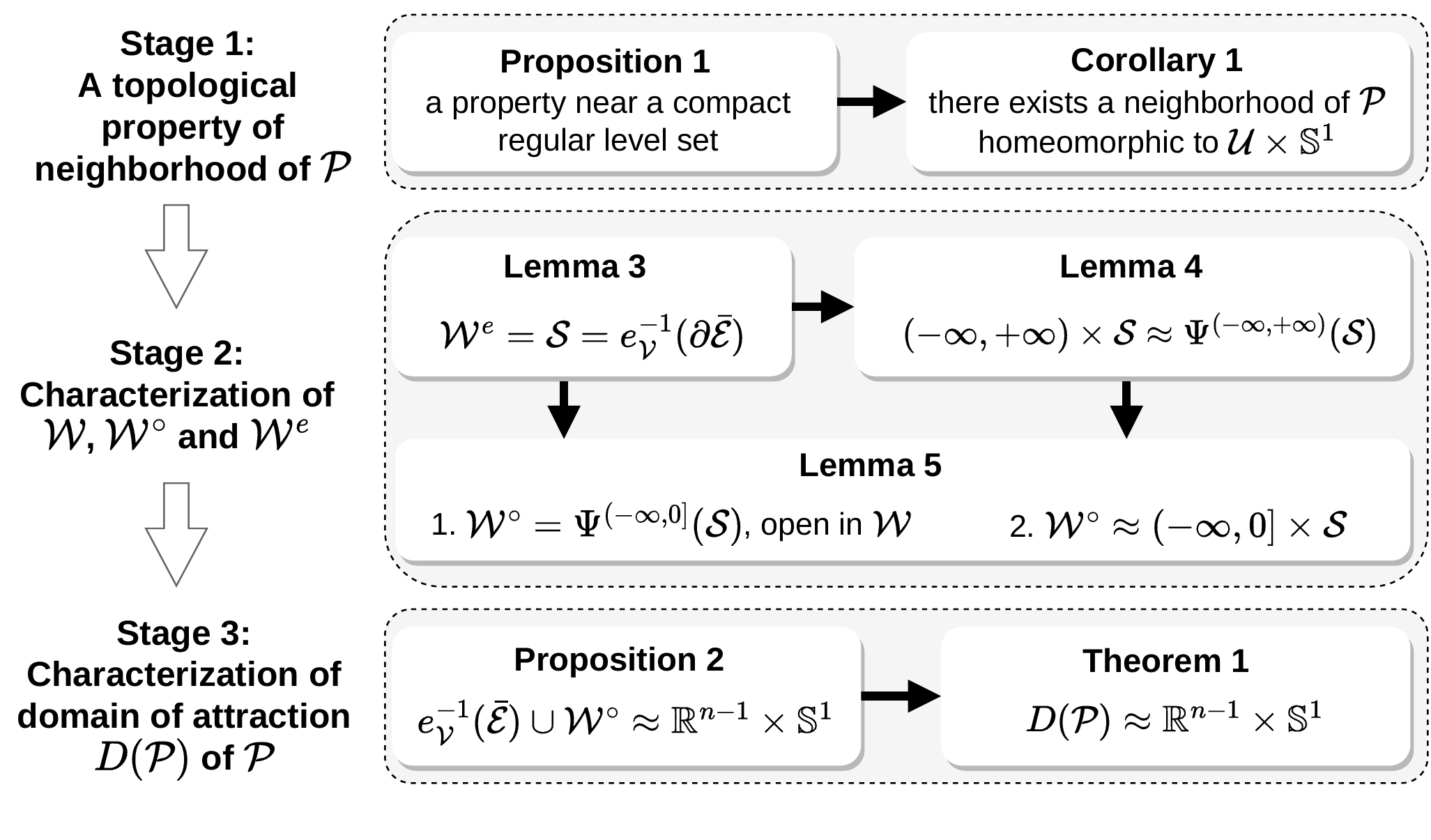}
	\caption{Overview and structure of the intermediate results leading to Theorem \ref{note_thm2}. Some conclusions of the propositions, lemmas and theorems are enclosed in the corresponding rounded rectangles.}
	\label{fig:overallproof}
\end{figure}

This subsection corresponds to Stage 1 in Fig. \ref{fig:overallproof}. There is a property holding near a \emph{compact component}\footnote{One can regard the level set itself as a topological space with the topology inherited from the ambient space, and then the concept of a \emph{component} of some level set is the same as that in the preliminaries (i.e., Section \ref{subsec_preliminary}).} of a \emph{regular} level set (i.e., the level set of a regular value of a smooth map). In particular, when the component of some regular level set of a smooth map is compact, then locally all level sets are compact and homeomorphic to the compact component. In addition, these level sets are placed ``nicely'' as precisely stated in the following proposition.

\begin{figure}[tb]
	\centering
	\includegraphics[width=0.7\linewidth]{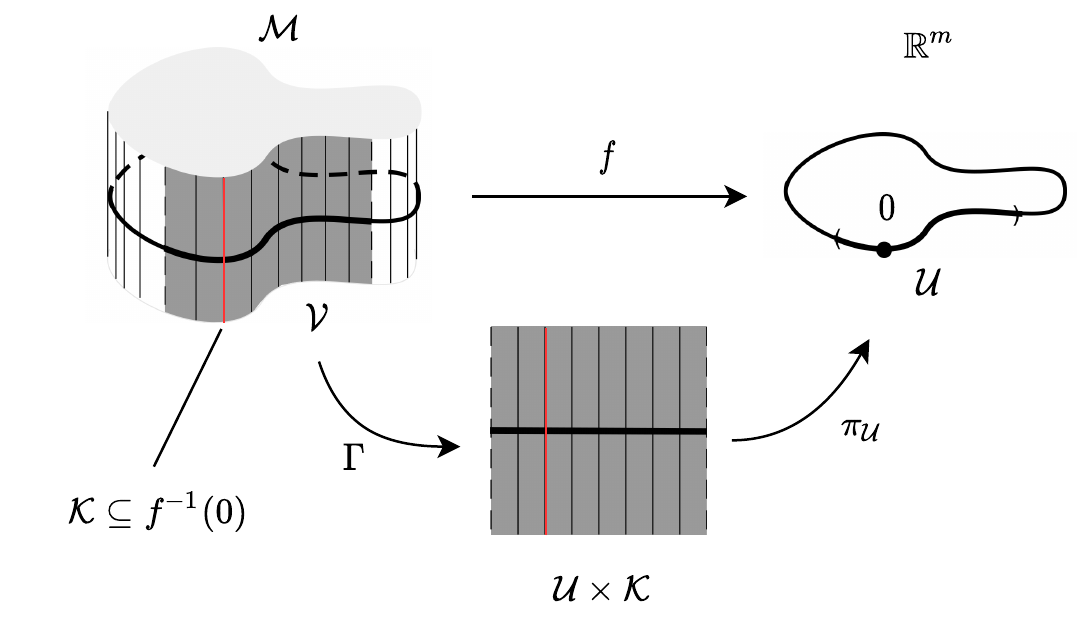}
	\caption{The illustration of Proposition \ref{note_prop_new}. The red vertical line in $\set{M}$ represents $\set{K}$, which is a compact component of the regular level set ${f}^{-1}(0)$, where $0 \in \set{U} \subseteq \mbr[m]$. The shaded region, denoted by $\set{V}$, is an open neighborhood  of $\set{K}$ in $\set{M}$, and $\set{U}$ is an open neighborhood of $0 \in \mbr[m]$.  All level sets in $\set{V}$ are topologically equivalent to $\set{K}$, and they are ``placed nicely'' in the sense that they  are homeomorphic to $\set{U} \times \set{K}$. This picture takes inspiration from \cite[Fig. 10.1]{lee2015introduction}.}
	\label{fig:localtrivializationdef2}
\end{figure}

\begin{prop}[Local triviality]\label{note_prop_new}
	Let $\set{M}$ be an $n$-dimensional smooth Riemannian manifold, and ${f}=(f_{1},...,f_{m}):\set{M}\rightarrow \mathbb{R}^{m}$ be a smooth map, where $m < n$. Suppose $0$ is a regular value of ${f}$ and $\set{K}\subseteq {f}^{-1}(0)$ is a compact component of ${f}^{-1}(0)$.  Then there exists an open neighbourhood $\set{V}$ of $\set{K}$ in $\set{M}$ with a diffeomorphism $\Gamma:\set{V} \rightarrow \set{U} \times \set{K} $ satisfying 
	$
	\pi_{\set{U}} \circ \Gamma = {f} |_{\set{V}},
	$
	where $\set{U} \subseteq \mbr[m]$ is an open neighborhood of $0$, $\pi_{\set{U}}$ denotes the projection of the product space $\set{U} \times \set{K} $ onto the first factor $\set{U}$, and ${f} |_{\set{V}}$ is the restriction of ${f}$ to $\set{V}$.
\end{prop}
The proof is shown in Appendix \ref{app1}. To explain Proposition \ref{note_prop_new}, for simplicity, we assume that there is only one component of ${f}^{-1}(0)$ and it is compact (i.e., $\set{K}={f}^{-1}(0)$). Then Proposition \ref{note_prop_new} claims that near the compact regular level set $\set{K}$ (i.e., in the neighborhood $\set{V}$), all level sets are homeomorphic to $\set{K}$. Moreover, these level sets are placed ``nicely'' in a way that is explicitly expressed by the homeomorphism $\Gamma$: they are homeomorphic to $\set{U} \times \set{K}  $ (see Fig. \ref{fig:localtrivializationdef2}). Roughly speaking, the level sets ${f}^{-1}(\set{U})$ in the open neighborhood $\set{V}$ are topologically equivalent to putting copies of $\set{K}$ in parallel and crossing them through by $\set{U}$, as illustrated by a thick line in Fig. \ref{fig:localtrivializationdef2} (i.e., $ \set{U} \times \set{K} $).  We can think of this property as showing the ``stability'' of the topology of level sets near the compact regular level set $\set{K}$. 

\begin{remark} \label{note_remark1}
	In this proposition, one can let the map ${f}$ be only \emph{twice continuously differentiable} (i.e., ${f} \in C^2$), and hence the map $\Gamma$ can be changed to a \emph{$C^2$-diffeomorphism}. 
\end{remark}

Let $\pi_i: \mbr[n] \to \mbr[]$ be the projection function that gives the $i$-th argument of a vector; namely, $\pi_i$ is defined by $\pi_i(x_1, \dots, x_i, \dots, x_n)=x_i$. Then one naturally has the following corollary related to the path-following problem.
\begin{coroll} \label{coroll1}
	There is an open neighborhood $\set{V}$ of the desired path $\set{P}=\inv{e}(0)$ in  $\manifold$ and a $C^2$-diffeomorphism $\Gamma$ from $\set{V}$ to $\set{U} \times \mathbb{S}^1 $, where $\set{U} = e(\set{V}) \subseteq \mbr[n-1]$, such that
	$
		e|_{\set{V}}=\pi_{\set{U}}\circ\Gamma,
	$
	where $\pi_{\set{U}}$ denotes the projection of the product space $\set{U} \times \mathbb{S}^1$ onto the first factor $\set{U} \subseteq \mbr[n-1]$ (i.e., projection of the first $(n-1)$ coordinates). Specifically, $e_{i}=\pi_{i} \circ \Gamma$ in $\set{V}$, for $i=1,\dots,n-1$.
\end{coroll}
\begin{proof}
	From Lemma \ref{lemmanifold}, the $C^2$ map $e = \transpose{({e}_1(\cdot), \dots, {e}_{n-1}(\cdot))}$ has a regular value $0 \in \mbr[n-1]$, and the inverse image is the desired path (i.e., $\set{P}:= e^{-1}(0)$), which is assumed to be homeomorphic to the unit circle (i.e., $\set{P} \approx \mathbb{S}^1$), hence compact \cite[Theorem 5.27]{lee2010topologicalmanifolds}. From Proposition \ref{note_prop_new} and Remark \ref{note_remark1}, the conclusions then follow.
\end{proof}
This implies that there exists a neighborhood $\set{U} \subseteq \mbr[n-1]$ of $0$ such that the preimage $\inv{e}(\set{U})$ ``looks like'' a solid torus $\set{U} \times \mathbb{S}^1$ in the homeomorphic sense. 


\subsection{Characterization of $\set{W}$, $\set{W}^{\circ}$ and $\set{W}^{e}$} \label{doa_path}
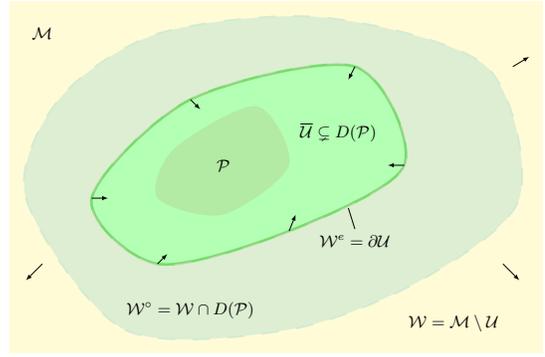
\begin{figure}[tb]
	\centering
	\begin{adjustbox}{width=0.8\columnwidth}
		\begin{tikzpicture}[long dash/.style={dash pattern=on 15pt off 15pt}]
			\begin{scope}[transparency group]
				\begin{scope}[blend mode=multiply]
					\fill[yellow!20]  (-9.5,-4.7) coordinate(A) rectangle (6.7,6) coordinate(B);
					\draw[cyan!40, ultra thick, long dash, fill=cyan!30, opacity=0.4] plot[smooth cycle] coordinates {(-9,0) (-6,-4) (-1,-4) (3,-2.5) (6,0) (5,4) (0,5.5) (-5,5) (-8,3)};
				\end{scope}
			\end{scope}
			
			\begin{scope}[transparency group]
				\begin{scope}[blend mode=multiply]
					\draw[green!30, line width=0.1cm, long dash, fill=green!30, opacity=1] plot[smooth cycle] coordinates {(-7,0) (-5,-2) (-1,-1) (2.5,1) (1,4) (-4,3)};
					\draw[cyan!40!yellow, line width=0.08cm, opacity=0.8] plot[smooth cycle] coordinates {(-7,0) (-5,-2) (-1,-1) (2.5,1) (1,4) (-4,3)};
					\fill[red!20, fill=red!20, opacity=0.4] plot[smooth cycle] coordinates {(-5,0) (-3,-0.5) (-1,1) (-2,2.8) (-4.5,1.5)};
				\end{scope}
			\end{scope}

			\draw [thick,-latex] plot [smooth, tension=1]  coordinates {(-7,0) (-6.5,0)};
			\draw [thick,-latex] plot [smooth, tension=1]  coordinates {(-5,-2) (-4.7,-1.7)};
			\draw [thick,-latex] plot [smooth, tension=1]  coordinates {(-1,-1) (-0.8,-0.5)};
			\draw [thick,-latex] plot [smooth, tension=1]  coordinates {(2.5,1) (2,1)};
			\draw [thick,-latex] plot [smooth, tension=1]  coordinates {(1,4) (0.8,3.6)};
			\draw [thick,-latex] plot [smooth, tension=1]  coordinates {(-4,3) (-3.7,2.7)};
			
			\draw [thick,-latex] plot [smooth, tension=1]  coordinates {(-8.5,-2) (-9,-2.5)};
			\draw [thick,-latex] plot [smooth, tension=1]  coordinates {(5.5,-2) (6,-2.5)};
			\draw [thick,-latex] plot [smooth, tension=1]  coordinates {(5.8,4) (6.3,4.3)};
			
			\node[scale=1.6](w) at (1.,-1.3) {$\set{W}^{e} = \partial \set{U}$};
			\node[scale=1.6] at (0.5,2) {$\overline{\set{U}} \subsetneq D(\set{P})$};
			\node[scale=1.6] at (-3,1) {$\set{P}$};
			\node[scale=1.6] at (-4,-3.4) {$\set{W}^{\circ} = \set{W} \cap D(\set{P})$};
			\node[scale=1.6] at (4,-3.8) {$\set{W} = \manifold \setminus \set{U}$};
			\node[scale=1.6] at (-8.5,5) {$\manifold$};
			\draw [thick] (w.north)-- (0.8, -0.3);
		\end{tikzpicture}
	\end{adjustbox}
	\caption{Illustration of $\set{W}$, $\set{W}^{\circ}$ and $\set{W}^{e}$ in $\manifold$. The red region (covered by green) is a compact asymptotically stable invariant set $\set{P}$, with the domain of attraction denoted by $D(\set{P})$. The green region represents an open and precompact neighborhood $\set{U}$ of $\set{P}$ and $\overline{\set{U}}$ is a proper subset of $D(\set{P})$. The yellow and cyan (covered by yellow) regions represent  $\set{W}$ and $\set{W}^{\circ}$ respectively. The black arrows are some vectors of a vector field on $\manifold$. All vectors on the boundary $\partial \set{U}$ are transverse to $\partial \set{U}$ and point to the interior of $\set{U}$. The exit set is $\set{W}^{e}=\partial \set{U}$, which acts like an exit of the set $\set{W}$.
	}
	\label{fig: eg_wazewski}
\end{figure}

Now we enter Stage 2 of the main results (see Fig. \ref{fig:overallproof}). We consider the autonomous system \eqref{eq1}. Let $t \mapsto \Psi(t,x_0)$ be the solution to \eqref{eq1} with the initial condition $\Psi(0,x_0)=x_0$, then $\Psi: \mbr[]_{\ge 0} \times \manifold \to \manifold$ is a flow \cite{chicone2006ordinary}. We will also use $\Psi^t(x_0)$ interchangeably with $\Psi(t,x_0)$, as is standard in the literature. Without loss of generality, we can assume that the solution to \eqref{eq1} is complete (i.e., the solution is well-defined for $t \in \mathbb{R}$), since otherwise one can replace the vector field $\vf$ by $\vf / (1+\norm{\vf}^2)$ without causing any difference to the topological properties of the flow that will be discussed in the sequel \cite[Proposition 1.14]{chicone2006ordinary}. Next, we define some sets, of which the definition is adapted from \cite[Def. 2.2]{conley1978isolated}.	%
\begin{defn}\label{note_def3}
	Given $\set{W}\subseteq\manifold$, called the \emph{pre-Wazewski set}, two subsets $\set{W}^{e}$ and $\set{W}^{\circ}$ of $\set{W}$ are defined to be
	$
		\set{W}^{e}	  \defeq \{{\xi}\in \set{W} : \forall t>0,  \Psi([0,t), {\xi})\not\subseteq \set{W} \}
	$
	and
	$
		\set{W}^{\circ} \defeq \{{\xi}\in \set{W} : \exists t>0, \Psi(t, {\xi}) \notin \set{W}\}, 
	$
	where $\not\subseteq$ means ``not a subset of''. The subsets $\set{W}^{e}$ and $\set{W}^{\circ}$ are called the \emph{exit set} and the \emph{pre-exit set} of the pre-Wazewski set $\set{W}$, respectively. 
\end{defn}
From the above definition, one observes that $\set{W}^{e} \subseteq \set{W}^{\circ} \subseteq \set{W}$. To understand these sets intuitively, an example is presented below.
\begin{example}
	Suppose, with abuse of notation, we have a compact asymptotically stable invariant set $\set{P} \subseteq \manifold$ (or a point when $\set{P}$ is a singleton), and its domain of attraction is denoted by $D(\set{P})$. We also assume that there is an open and precompact neighborhood $\set{U} \supseteq \set{P}$ of $\set{P}$ such that i) The closure of $\set{U}$, denoted by $\overline{\set{U}}$, is a (proper) subset of the domain of attraction $D(\set{P})$; ii) every trajectory starting from the boundary $\partial \set{U}$ of $\set{U}$ immediately leaves the boundary and enters the neighborhood $\set{U}$ (and thus converges to $\set{P}$ subsequently). Therefore, every trajectory starting from $\set{U}$ converges to $\set{P}$.  Define the pre-Wazewski set as $\set{W}=\manifold \setminus \set{U}$, which is a closed set in $\manifold$. Then the aforementioned boundary $\partial \set{U} \subseteq \set{W}$ is the exit set $\set{W}^{e}$, since every trajectory starting from it immediately leaves the set $\set{W}$, and thus it acts like an \emph{exit} of $\set{W}$. Since $\set{U}$ is a proper subset of the domain of attraction $D(\set{P})$, we define the pre-exit set as $\set{W}^{\circ} = \set{W} \cap D(\set{P}) \ne \emptyset$. The pre-exit set $\set{W}^{\circ}$ is such that every trajectory starting from $\set{W}^{\circ}$ may stay in $\set{W}$ for some time (in contrast to $\set{W}^{e}$ for which trajectories leave $\set{W}$ immediately) but leaves $\set{W}$ and enters $\set{U}$ eventually (Fig. \ref{fig: eg_wazewski}). 
	
	Through this example, the pre-exit set $\set{W}^{\circ}$ and the exit set $\set{W}^{e}$ may be seen as a generalization of some neighborhoods of $\set{P}$, while $\set{W}$ itself is a generalization or, better expressed, variation on the set of points in the \emph{exterior} of the domain of attraction of $\set{P}$, but also actually including part of that domain of attraction. 
\end{example}

We can find the sets $\set{W}, \set{W}^{\circ}, \set{W}^{e}$, and then characterize the domain of attraction by taking advantage of the physical intuition associated with these sets. %
For simplicity, we denote by $e_{\set{V}}$ the restriction of $e$ to $\set{V}$ (i.e., $e_\set{V} \defeq e|_\set{V}$), where $\set{V}$ is defined in Corollary \ref{coroll1}. In addition, we define an ellipsoid:
\begin{equation} \label{note_eqD}
	\set{E} \defeq \{ x \in \mbr[n-1] : \transpose{x} K x < R \} \subseteq \mbr[n-1]
\end{equation}
centered at $0 \in \set{U}=e(\set{V}) \subseteq \mbr[n-1]$, where $K \defeq \diag{k_1, \dots, k_{n-1}}$ is the diagonal matrix with all the positive gains $k_i$, $i=1,\dots,n-1$, and $R>0$ is chosen sufficiently small such that $\overline{\set{E}}\subseteq \set{U}$, where $\overline{\set{E}}$ denotes the closure of $\set{E}$ and $\set{U}$ is defined in Corollary \ref{coroll1}. Since $e_{\set{V}}^{-1}(\overline{\set{E}}) \approx \overline{\set{E}}\times \mathbb{S}^1$ by Corollary \ref{coroll1}, $e_{\set{V}}^{-1}(\overline{\set{E}})$ is an embedded submanifold with the manifold boundary \cite[p. 120]{lee2015introduction}
\begin{equation} \label{note_eqs}
	\set{S} \defeq e_{\set{V}}^{-1}(\partial\overline{\set{E}}) \subseteq \set{M},
\end{equation}
which is homeomorphic to $\partial\overline{\set{E}}\times \mathbb{S}^1$.	We define the pre-Wazewski set $\set{W}$ below:
\begin{equation} \label{note_eqw}
	\set{W} \defeq \manifold \setminus e_{\set{V}}^{-1}(\set{E}) \subseteq \set{M}.
\end{equation}
With the introduction of the exit set in Definition \ref{note_def3}, the next lemma identifies exactly what the exit set of $\set{W}$ is, and also shows the forward-invariance of $e_{\set{V}}^{-1}(\overline{\set{E}}) \subseteq \set{M}$ and $e_{\set{V}}^{-1}(\set{E}) \subseteq \set{M}$.
\begin{lemma} \label{note_lemma3}
	There holds that
	\begin{enumerate}[label=(\roman*)]
		\item \label{claim3.1} The exit set $\set{W}^{e} = \set{S}$;
		\item \label{claim3.2} The sets $e_{\set{V}}^{-1}(\overline{\set{E}})$ and $e_{\set{V}}^{-1}(\set{E})$ are forward-invariant;
		\item \label{claim3.3} For the initial condition $\xi(0) \in e_{\set{V}}^{-1}(\overline{\set{E}})$, the trajectory $\xi(t)$ of \eqref{eq1} converges asymptotically to the desired path as $t \to \infty$;
		\item \label{claim3.4} The set $\Psi^{(0,+\infty)}(\set{S})\subseteq e_{\set{V}}^{-1}(\set{E})$, where $\Psi^{(0,+\infty)}(\set{S}) \defeq \{\Psi^{t}({\xi}) : t \in (0, +\infty), {\xi} \in \set{S} \}$.
	\end{enumerate}
\end{lemma}
The main idea of the proof (see Appendix \ref{app2}) is to use the Lyapunov function $e(\cdot)^\top K e (\cdot)$ to study the properties of the vector field defined on the topologically ``clearer'' space $\set{U} \times \mathbb{S}^1$. This lemma identifies the exit set and invariant sets, which will be useful for the proofs of Lemmas \ref{note_lemma4} and \ref{note_lemma5}. The following lemma reveals that the flow $\Psi(\cdot,\cdot)$ of \eqref{eq1} is a homeomorphism.
\begin{lemma} \label{note_lemma4}
	The flow $\Psi : \mbr[] \times \set{S}  \to \manifold$ of \eqref{eq1} is an open injection. In particular, $\Psi : (-\infty,+\infty) \times \set{S} \to \Psi^{(-\infty,+\infty)}(\set{S})$ is a homeomorphism.
\end{lemma}
The proof in Appendix \ref{app3} follows standard topological arguments. This lemma identifies homeomorphic objects related by the homeomorphism $\Psi^{(\cdot)}(\cdot)$. By Definition \ref{note_def3}, the pre-exit set $\set{W}^{\circ}$ is particularized to
\begin{equation} \label{eqwcirc}
	\set{W}^{\circ}=\{x \in \set{W} : \exists t>0, \Psi^{t}(x)\in e_{\set{V}}^{-1}(\set{E})\}.
\end{equation}
Based on Lemmas \ref{note_lemma3} and \ref{note_lemma4}, we can give the exact expression of $\set{W}^{\circ}$.
\begin{lemma} \label{note_lemma5}
	There holds that
	\begin{enumerate}[label=(\roman*)]
		\item \label{claim5.1} The pre-exit set $\set{W}^{\circ}=\Psi^{(-\infty,0]}(\set{S})$, and it is open in $\set{W}$. 
		\item \label{claim5.2} The pre-exit set $\set{W}^{\circ}$ is homeomorphic to $(-\infty,0] \times \set{S}$ given by the homeomorphism $\Psi^{(\cdot)}(\cdot)$.
	\end{enumerate}
\end{lemma}
The proof (see Appendix \ref{app4}) is derived directly from the properties/definitions of the exit set, invariant sets and the homeomorphism $\Psi^{(\cdot)}(\cdot)$ in Lemmas \ref{note_lemma3} and \ref{note_lemma4}. 

\subsection{Domain of attraction of the desired path} \label{doa_path2}
\begin{figure}
	\centering
	\includegraphics[width=0.4\linewidth]{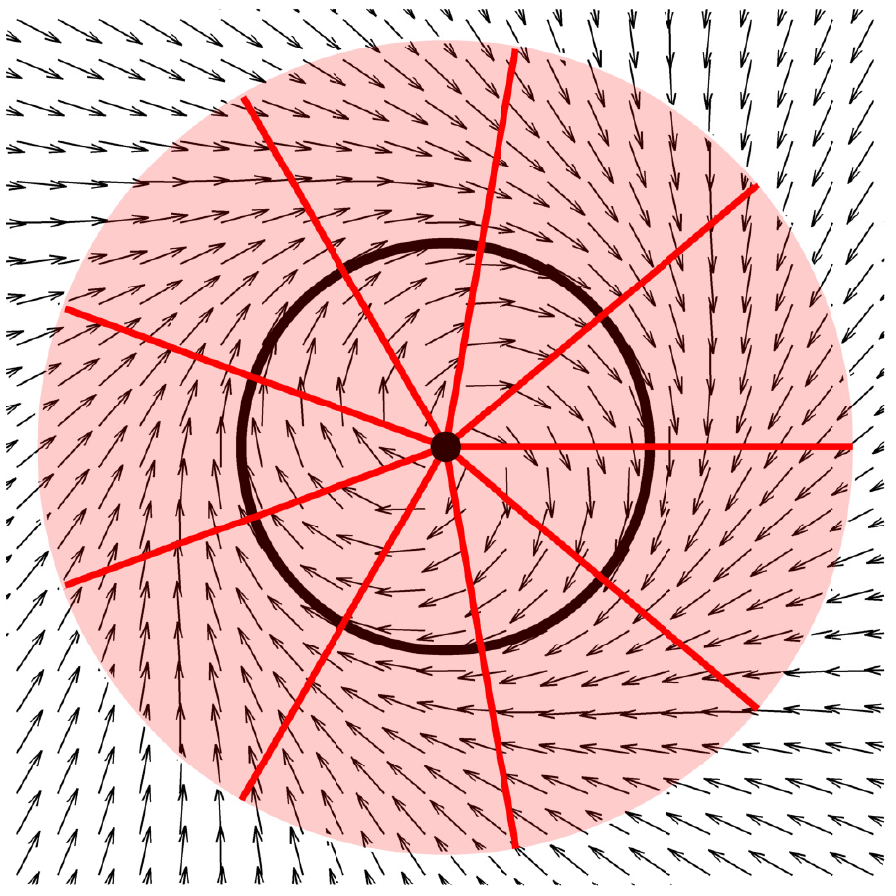}
	\caption{The black solid curve is the desired path $\set{P}$ and the black dot is the singular point. The red solid radiant lines are homeomorphic to $\mathbb{R} \times \{p\}$, where $p \in \set{P}$. The domain of attraction of $\set{P}$ is homeomorphic to $\cup_{p \in \set{P}} \, \mathbb{R} \times \{p\} \approx \mathbb{R} \times \mathbb{S}^1$, and resembles the red transparent disk. }
	\label{fig:circlevf}
\end{figure}

We can now characterize the domain of attraction of the desired path (i.e., Stage 3 of Fig. \ref{fig:overallproof}),  which is related to the ``shape'' of the set $e_{\set{V}}^{-1}(\overline{\set{E}})\cup \set{W}^{\circ}$.
\begin{prop} \label{note_prop3}
	The set $e_{\set{V}}^{-1}(\overline{\set{E}})\cup \set{W}^{\circ}$ is homeomorphic to $\mbr[n-1]\times \mathbb{S}^1$.
\end{prop}
The proof is shown in Appendix \ref{app5}, which relies on Corollary \ref{coroll1}, the homeomorphism $\Psi^{(\cdot)}(\cdot)$, and the construction of quotient maps. In the context of the vector-field guided path-following problem, Proposition \ref{note_prop3} leads to the following theorem.
\begin{theorem} \label{note_thm2}
	Suppose the desired path $\set{P}$ is asymptotically stable and homeomorphic to the unit circle $\mathbb{S}^1$, then the set of initial conditions such that trajectories of \eqref{eq1} asymptotically converge to the desired path $\set{P}$ is homeomorphic to $\mbr[n-1] \times \mathbb{S}^1$. 
\end{theorem}
\begin{proof}
	One observes that $e_{\set{V}}^{-1}(\overline{\set{E}})\cup \set{W}^{\circ}$ consists of initial conditions such that trajectories of \eqref{eq1} asymptotically converge to the desired path $\set{P}$ as $t \to \infty$. The result then follows from Proposition \ref{note_prop3}.
\end{proof}
An example below demonstrates the significance of this theorem. 
\begin{example}
	Suppose the desired path is the circle described by $\set{P}=\{(x,y) \in \mathbb{R}^2: e(x,y)=0 \}$, where $e(x,y)=x^2+y^2-4$. Therefore, the vector field is $\vf=E \gradient e - k\, e \gradient e$, where $E = \left[ \begin{smallmatrix} 0 & -1  \\ 1 & 0 \end{smallmatrix} \right]$ is the $90^\circ$ rotation matrix and $k$ is a positive gain. There is only one singular point (i.e., the origin). According to \cite[Theorem 2]{yao2021topo}, the domain of attraction $D(\set{P})$ is homotopy equivalent to $\mathbb{S}^1$, which implies that $D(\set{P})$ may be, e.g., $\set{P}$ itself, or $\mathbb{R}^2 \setminus \{0\}$, or roughly speaking, a circle with segments attached (shapes like ``$9$'', ``$6$''). If $D(\set{P})=\set{P}$, this implies that the desired path is \emph{non-attractive}, which is incorrect according to \cite[Corollary 1]{yao2021topo}. Therefore, \cite[Theorem 2]{yao2021topo} is insufficient to provide a satisfactory answer, or may even lead to incorrect conclusions. However, according to the strengthened result, Theorem \ref{note_thm2}, the possibility of $D(\set{P})=\set{P}$ is excluded, since $\set{P}$ is \emph{not} homeomorphic to $\mathbb{R} \times \mathbb{S}^1$, and thus implies that $\set{P}$ is \emph{not} non-attractive. The strengthened result also implies that $D(\set{P})$ cannot be a circle with segments attached (shapes like ``$9$'', ``$6$''). In addition, Theorem \ref{note_thm2} provides a more intuitive impression on the shape of the domain of attraction, which is similar to an unbounded disk consisting of an infinite number of rays radiating from the origin (see Fig. \ref{fig:circlevf}). From Theorem \ref{note_thm2}, one can also conclude that global convergence to the desired path is impossible, since $\mathbb{R}^2$ is \emph{not} homeomorphic to $\mathbb{R} \times \mathbb{S}^1$. In fact, the domain of attraction is $\mathbb{R}^2 \setminus \{0\}$ \cite{kapitanyuk2017guiding}. Another implication of Theorem \ref{note_thm2} is that, given the system \eqref{eq1}, if global convergence to the desired path is required, then the only feasible solution is to change the topology of the desired path in order to change the topology of the domain of attraction \cite{yao2020singularity}. In $\mathbb{R}^3$, one can also conclude from Theorem \ref{note_thm2} that the domain of attraction \emph{cannot} be a M\"{o}bius strip.  
\end{example}
Using the same idea of the proof, we can easily extend Theorem \ref{note_thm2} to a $k$-dimensional compact manifold with the following modifications of the problem formulation:
(i) The definition of $\set{P}$ is changed to $\set{K} = \{ \xi \in \manifold: e_i(\xi)=0, i=1,\dots,n-k \}$, and $\set{K}$ is assumed to be a compact regular zero-level set of $e(\cdot)=(e_1(\cdot),\dots,e_{n-k}(\cdot))^{\top}$; (ii) The definition of the vector field \eqref{new_gvf} is properly modified such that the term $\bot_e$ is well defined and orthogonal to the $\gradient e_i$ terms, for $i=1,\dots,n-k$. The new definition of $\bot_e$ is elaborated in \cite[Section IV.B]{yao2022multitro} and \cite[Section VI.D]{yao2021collision}.
\begin{theorem}[Generalization of Theorem \ref{note_thm2}]
	If $\set{K}$ is a $k$-dimensional compact asymptotically stable manifold, where $k \ge 1$, then the domain of attraction of $\set{K}$ is homeomorphic to $\mathbb{R}^{n-k} \times \set{K}$.
\end{theorem}
\begin{proof}
	The idea of the proof is the same as that of Theorem \ref{note_thm2}. 
\end{proof}

\begin{remark} \label{remark_wilson}
	In Theorem 3.4 of \cite{wilson1967structure}, it is claimed that the domain of attraction of a compact or non-compact \emph{uniformly asymptotically stable}\footnote{Let $\xi(t)$ denote a trajectory of an autonomous system. In \cite{wilson1967structure}, a closed set $\set{A}$ is said to be \emph{uniformly asymptotically stable} if there exists $r>0$ such that for every $\epsilon>0$, there exists a time instant $T(\epsilon)$ such that $\dist(\xi(0), \set{A})<r \implies \dist(\xi(t), \set{A}) < \epsilon$ for $t > T(\epsilon)$. Uniform asymptotic stability is stronger than \emph{asymptotic stability}, but if $\set{A}$ is compact, then they are equivalent. } submanifold $\set{A}$ of a finite-dimensional manifold  $\manifold$ is diffeomorphic to an open tubular neighborhood of $\set{A}$. Therefore, Theorem \ref{note_thm2} is consistent with this claim. However, %
	the strong claim of \cite[Theorem 3.4]{wilson1967structure} without imposing the compactness requirement on $\set{A}$ is not accurate, as shown in Counterexample \ref{counterexample1} below. Our work \cite{lin2021wilson} develops a detailed proof of \cite[Theorem 3.4]{wilson1967structure} with an additional assumption that $\set{A}$ is compact.%
\end{remark}
\begin{counterexample} \label{counterexample1}
	Consider the manifold 
	$
	\manifold=\mbr[2] \setminus \{(-1,0), (1,0)\}
	$
	with the subspace topology inherited from $\mbr[2]$. Define 
	$
	\set{A} \defeq (-1,1) \times \{0\} \subseteq \manifold
	$
	as a subspace of $\set{M}$. One can verify that $\set{A}$ is closed in $\manifold$ and is a submanifold of $\manifold$. Note that $\set{A}$ is \emph{not} compact since the open cover $\set{U}_\epsilon \defeq \{(-1+\epsilon, 1-\epsilon) \times (-\epsilon, \epsilon) : \forall \epsilon>0 \}$ of $\set{A}$ does not have a finite subcover. Define $f: \manifold \to \mbr[]_{\ge 0}$ to be
	\begin{align*}
		\scalemath{0.9}{	
		f({\xi}) = \left( \dist({\xi}, \set{A}) \right)^2 
		= 
		\begin{dcases}  
			y^2, & \text{ if } -1<x<1 \\
			(x-1)^2+y^2, & \text{ if } x>1 \text{ or } x=1, y \ne 0 \\
			(x+1)^2+y^2, & \text{ if } x<-1 \text{ or } x=-1, y \ne 0,
		\end{dcases}
	}
	\end{align*}
	where ${\xi}=(x,y) \in \set{M}$. One can verify that $f(\cdot)$ is continuously differentiable. Indeed, the partial derivatives of $f$ w.r.t. $x$ and $y$ are $\frac{\partial f}{\partial x}=0$, if $-1<x<-1$ or $x=\pm 1, y \ne 0$; $\frac{\partial f}{\partial x}=2(x-1)$, if $x >1$; $\frac{\partial f}{\partial x}=2(x+1)$, if $x<1$,
	and $\frac{\partial f}{\partial y} = 2y$, respectively. Therefore, the gradient system
	$
		\dot{{\xi}} = - \gradient f({\xi}),
	$
	renders $\set{A}$ globally uniformly asymptotically stable (proved by using the radially unbounded Lyapunov function $V({\xi})=f^2({\xi})$). Specifically, the domain of attraction of $\set{A}$ is the manifold $\manifold$, which is \emph{not} contractible. However, according to \cite[Theorem 3.4]{wilson1967structure}, the domain of attraction of $\set{A}$ is homeomorphic to its tubular neighborhood, but its tubular neighborhood is homeomorphic to $\set{A} \times \mbr[]$, which is contractible. Therefore, this counterexample shows that the claim of \cite[Theorem 3.4]{wilson1967structure} without requiring the compactness of $\set{A}$ is inaccurate. Furthermore, the domain of attraction of $\set{A}$ is not even homotopy equivalent to $\set{A}$ since the fundamental group of the domain of attraction, i.e., $\manifold$, is non-zero, but that of $\set{A}$ is $0$ as it is contractible. 
\end{counterexample}

\section{Conclusion and future fork}  \label{sec_conclusion}
It has been shown that the domain of attraction of a desired path $\set{P}$, which is a compact asymptotically stable one-dimensional embedded submanifold under an autonomous system, is homotopy equivalent to the unit circle $\mathbb{S}^1$ \cite{yao2021topo,moulay2010topological}. However, homotopy equivalent objects may be geometrically distinctive, so this characterization of the domain of attraction sometimes may not be satisfactory. In this paper, we strengthen this result and show that the domain of attraction is homeomorphic to $\mbr[n-1] \times \mathbb{S}^1$, where $n$ is the dimension of the ambient manifold $\manifold$. We also provide an example showing that if the considered submanifold is not compact (see Remark \ref{remark_wilson}), then its domain of attraction is \emph{not} even homotopy equivalent to the submanifold, not to mention Theorem \ref{note_thm2}. An undergoing study \cite{lin2021wilson} is to provide a detailed proof of a refined theorem of \cite[Theorem 3.4]{wilson1967structure}. One may hypothesize that if the desired path is \emph{diffeomorphic} to $\mathbb{S}^1$, then the domain of attraction is \emph{diffeomorphic} to $\mathbb{R}^{n-1} \times \mathbb{S}^1$. The investigation of this hypothesis is left for future work.

\appendix
\subsection{Proof of Proposition \ref{note_prop_new}} \label{app1}
	Proposition \ref{note_prop_new} may be proved using the Ehresmann theorem \cite[p. 378]{cushman1997global} or \cite[Proposition 2.1]{lukina2008global}. Alternatively, we provide a self-contained proof to give more insights. The main idea is letting the neighborhood $\set{V}$ in Proposition \ref{note_prop_new} be a subset of the tubular neighbourhood $\mathrm{Tub}$ of $\set{K}$, which is diffeomorphic to the normal bundle of $\set{K}$ \cite[Proposition 7.1.3]{mukherjee2015differential}, and $\Gamma$ is taken ``similar to'' $({f}, r)$, where $r:\mathrm{Tub} \to \set{K}$ is a retraction (note that in the proof, we do not construct the retraction $r$). Roughly speaking, the map $H$ shown in the proof is essentially the map $({f},r)$, except that its domain is $\mathbb{R}^{m} \times \set{K}$ rather than $\set{M}$ (see \textbf{Step 2}). To change the domain from $\mathbb{R}^{m} \times \set{K}$ to $\set{M}$ (see \textbf{Step 3}), we need to utilize the triviality of the normal bundle (see \textbf{Step 1}). These three steps of the proof are illustrated below.
\begin{proof}[Proof of Proposition \ref{note_prop_new}]	
	Under the assumptions of Proposition \ref{note_prop_new},  $\set{K}$ consists of regular points of ${f}$, and it is an $(n-m)$-dimensional embedded submanifold of $\set{M}$ \cite[Corollary 5.14]{lee2015introduction}. Since $\set{K}$ is open in ${f}^{-1}(0)$ w.r.t. the subspace topology\footnote{As $0$ is a regular value, ${f}^{-1}(0)$ is an embedded submanifold of $\manifold$, and thus it is locally connected. By \cite[Theorem 25.3]{munkres2013topology}, since $\set{K}$ is a component of ${f}^{-1}(0)$, it is open (in addition to being closed) in ${f}^{-1}(0)$. }, there exists an open neighbourhood $\set{O}$ in $\set{M}$ such that $\set{O} \cap {f}^{-1}(0) = \set{K}$ (hence, ${f}|_{\set{O}}^{-1}(0) = \set{K}$), and ${f}|_{\set{O}}$ is a submersion. Therefore, without loss of generality, we can  simply assume henceforth that ${f}$ is a submersion for all points in a neighborhood of $\set{K}$ and $\set{K}={f}^{-1}(0)$. 
	
	The proof can be decomposed into the following steps.
	
	\textbf{Step 1 (Construct the map $\Theta$ via the normal bundle to relate $\mathbb{R}^{m} \times \set{K}$ and $\manifold$):} We show in the following lemma that the normal bundle $N_{\set{K}} = \sqcup_{{\xi} \in \set{K}} N_{\xi}$ of $\set{K}$  is trivial (i.e., $N_{\set{K}}$ is isomorphic to $\mathbb{R}^{m} \times \set{K}$), where $\sqcup$ is the disjoint union.
	\begin{lemma} \label{lemma_trivialnormalbundle}
		The normal bundle $N_{\set{K}}$ of $\set{K}$ in $\set{M}$ is trivial. Namely, $N_{\set{K}}\cong\mathbb{R}^{m} \times \set{K}$.
	\end{lemma}
	\begin{proof}[Proof of Lemma \ref{lemma_trivialnormalbundle}]
		For each ${\xi}\in\set{K}$, since $\set{K}$ consists of regular points, the gradient vectors $\gradient f_{1}({\xi}), \dots, \gradient f_{m}({\xi})$ constitute a basis of the normal vector space $N_{{\xi}} \defeq \text{span}\{\gradient f_{1}({\xi}), \dots, \gradient f_{m}({\xi})\}$ with $N_{{\xi}}\subseteq T_{{\xi}}\set{M}$ and $N_{{\xi}}\perp T_{{\xi}}\set{K}$. Since $\gradient f_{i}({\xi})$, $i=1,\dots,m$, vary smoothly w.r.t. ${\xi}$, $\{ \gradient f_{1},...,\gradient f_{m} \}$ constitutes a smooth global frame for the normal bundle $N_{\set{K}} = \sqcup_{{\xi} \in \set{K}} N_{\xi}$ of $\set{K}$. Therefore, $N_{\set{K}}$ is a trivial vector bundle; i.e., $N_{\set{K}}\cong\mathbb{R}^{m} \times \set{K}$ \cite[Corollary 10.20]{lee2015introduction}.
	\end{proof}
	As a result of the above lemma, there is an open embedding\footnote{The reason why the domain is $\mathbb{R}^{m} \times \set{K} $ rather than $\set{D} \times \set{K}$, for some open neighborhood $\set{D}$ in $\mathbb{R}^m$, is that, in the map $\Theta$, there is the homeomorphism $\rho$ mapping an open ball of radius $\epsilon>0$ (a uniform radius exists due to the compactness of $\set{K}$) to $\mathbb{R}^m$; i.e., $\rho(v) = v / (\epsilon -\|v\|)$, for $v \in \set{B}_\epsilon \subseteq \mathbb{R}^m$, where $\set{B}_\epsilon$ is the open ball with radius $\epsilon$, and the inverse map is $\rho^{-1}(w)= \epsilon w / (1+\|w\|)$ for $w \in \mathbb{R}^m$ \cite[Theorem 7.1.5, Proposition 7.1.3]{mukherjee2015differential}. }  
	$
	\Theta: \mathbb{R}^{m} \times \set{K}\rightarrow \set{M}
	$
	such that $\Theta(0, {\xi}) = {\xi}$ for any ${\xi} \in \set{K}$. 
	\begin{figure}[tb]
		\centering
		\begin{tikzpicture}
			\matrix (m) [matrix of math nodes,row sep=3em,column sep=3em,minimum width=2em]
			{ \mathbb{R}^{m} \times \set{K} & \manifold & \mathbb{R}^{m} \\
			\mathbb{R}^{m} \times \set{K} & {} & \mathbb{R}^{m} \times \set{K}  \\
			\set{V} \subseteq \manifold & \set{B}_{\epsilon} \times \set{K} &  \set{U} \times \set{K} \\ };
			\path[-stealth]
			(m-1-1) edge node [above] {$\Theta$} (m-1-2)
			(m-1-2) edge node [above] {$f$} (m-1-3);
			\draw [->, >=stealth, out=-20, in=-160, looseness=0.55] (m-1-1.south) to node[below]{$g = f \circ \Theta$} (m-1-3.south);
			\path[-stealth] (m-2-1) edge node [below] {$h=g \times \identity_{\set{K}}$} (m-2-3);
			\path[-stealth]
			(m-3-1) edge node [above] {$\Theta^{-1}$} (m-3-2) 
			(m-3-2) edge node [above] {$h_{\epsilon}$} (m-3-3);
			\draw [->, >=stealth, out=-20, in=-160, looseness=0.55] (m-3-1.south) to node[below]{$\Gamma = h_{\epsilon} \circ \Theta^{-1}$} (m-3-3.south);
		\end{tikzpicture}
		\caption{The construction of the map $\Gamma = h_{\epsilon} \circ \Theta^{-1}$ (in \textbf{Step 3}) involves the maps $\Theta$ (in \textbf{Step 1}), $g= f \circ \Theta$ and $h = g \times \identity_{\set{K}}$ (in \textbf{Step 2}). }
		\label{fig:gh gamma maps}
	\end{figure}
	
	\textbf{Step 2 (Construct the map $h$ to relate $\mathbb{R}^{m} \times \set{K}$ and $\mathbb{R}^{m} \times \set{K}$, approximating $\Gamma$):} By restricting ${f}$ to the tubular neighborhood $\text{Im}\,\Theta$, where $\text{Im}$ denotes the image of a map, we can turn our focus from the map ${f}$ in Proposition \ref{note_prop_new} to the following map 
	$
	g = {f} \circ \Theta:\mathbb{R}^{m} \times \set{K}\rightarrow\mathbb{R}^{m},
	$
	which is a submersion in the vicinity of $g^{-1}(0)=\{0\} \times \set{K}$. Therefore, we can define a new map $h : \mathbb{R}^{m} \times \set{K} \to \mathbb{R}^{m} \times \set{K}$ by $h(v, {\xi}) =\big(g(v,{\xi}), {\xi} \big).$
	Observe that the map $h$ is a local diffeomorphism in a neighborhood of $\{0\} \times \set{K}$, since one can check that its tangent map at each point in a neighborhood of $\{0\} \times \set{K}$ is an isomorphism. 
	
	Now we show that with $\epsilon>0$ sufficiently small, the restriction of $h$ to $ \set{B}_{\epsilon} \times \set{K}$, denoted by $h_{\epsilon}$ (i.e., $ h_{\epsilon} = h |_{ \set{B}_{\epsilon} \times \set{K}}$), is injective. To this end, we only need to show that $g$ is injective on $\set{B}_{\epsilon} \times \{{\xi}\}$ for every ${\xi}\in\set{K}$. This is simply because if $ h(\bar{v},\bar{{\xi}})=h(v,{\xi})$, then $\bar{{\xi}}={\xi}$, and then $g$ being injective on $ \set{B}_{\epsilon} \times \{{\xi}\} $ implies $v=\bar{v}$. The injectivity of $g$ is formally stated in the following lemma.
	\begin{lemma} \label{lemma_injective}
		With $\epsilon>0$ sufficiently small, $g$ is injective on $ \set{B}_{\epsilon} \times \{{\xi}\}$ for every ${\xi}\in\set{K}$.
	\end{lemma}
	\begin{proof}[Proof of Lemma \ref{lemma_injective}]
		As mentioned above, $h:\mathbb{R}^{m} \times \set{K}\rightarrow\mathbb{R}^{m} \times \set{K}$ is a local diffeomorphism in a neighborhood of $\{0\} \times \set{K}$. Therefore, for each $p\in\set{K}$, there exists an open neighbourhood $\set{U}_{p}$ in $\set{K}$ and $\epsilon_{p}>0$ such that $h$ is injective on $ \set{B}_{\epsilon_{p}} \times \set{U}_{p} $. In particular, $h$ is injective on $ \set{B}_{\epsilon_{p}} \times \{q\} $ for every $q\in \set{U}_{p}$. Since $\set{K}$ is compact and the family of open neighborhoods $\{\set{U}_{p} : p\in\set{K}\}$ covers $\set{K}$, there exists a finite subcover $\{\set{U}_{p_{1}}, \dots ,\set{U}_{p_{a}}\}$ of $\set{K}$. Let $\epsilon=\min\{\epsilon_{p_{1}},...,\epsilon_{p_{a}}\}$. For each ${\xi}\in\set{K}$, we have ${\xi}\in \set{U}_{p_{i}}$ for some $i\in\{1,...,a\}$, and $h$ is then injective on $ \set{B}_{\epsilon} \times \{{\xi}\} $ for every ${\xi} \in \set{K}$.
	\end{proof}
	
	\textbf{Step 3 (Construct the map $\Gamma$ by changing the domain of $h$):} The injectivity of $h_{\epsilon} $ implies that
	$
	h_{\epsilon}: \set{B}_{\epsilon} \times \set{K}\rightarrow h( \set{B}_{\epsilon} \times \set{K})
	$
	is a diffeomorphism with $h( \set{B}_{\epsilon} \times \set{K})$ being an open neighbourhood of $\{0\} \times \set{K}$ in $\mathbb{R}^{m} \times \set{K}$. Since $\set{K}$ is compact, there exists $\delta>0$, such that $ \set{B}_{\delta} \times \set{K} $ is contained in $h( \set{B}_{\epsilon} \times \set{K})$ \cite[Lemma 26.8]{munkres2013topology}. Let $\set{U} \defeq \set{B}_{\delta}$ and $\set{V} \defeq \Theta \circ h_{\epsilon}^{-1}( \set{U} \times \set{K} )$, we have
	$
	\Gamma=h_{\epsilon}\circ\Theta^{-1}:\set{V} \to \set{U} \times \set{K} 
	$
	which is a diffeomorphism satisfying $\pi_{\set{U}}\circ\Gamma={f}$ (see Fig. \ref{fig:gh gamma maps}). 
\end{proof}

\subsection{Proof of Lemma \ref{note_lemma3}} \label{app2}
\begin{proof}
	\textbf{Proof of Claim \ref{claim3.1}:} First note that
	\begin{multline} \label{eq_ortho_zero}
		\langle \gradient {e}_i, \vf \rangle_g \stackrel{\eqref{new_gvf}}{=} 
		\left\langle \gradient {e}_i, \; \orthoterm - \sum_{j=1}^{n-1} k_j {e}_j \gradient {e}_j \right\rangle_g \\
		=  \left\langle \gradient {e}_i, - \sum_{j=1}^{n-1} k_j {e}_j \gradient {e}_j \right\rangle_g,
	\end{multline}
	for $i=1,\dots,n-1$, where $\langle \cdot, \cdot \rangle_g$ denotes the Riemannian metric, and the orthogonality property (Lemma \ref{lemma1}) is used in the last equation. One can calculate the time derivative of the path-following error $e$:
	\begin{equation} \label{eq_e_dot}
		\begin{split} 
			\dt e(\xi(t)) &= \dt \matr{{e}_1(\xi(t)) \\ \vdots \\ {e}_{n-1}(\xi(t))} = \matr{\langle \gradient {e}_1, \vf \rangle_g \\ \vdots \\ \langle \gradient {e}_{n-1}, \vf \rangle_g} \\
			&\stackrel{\eqref{eq_ortho_zero}}{=} \matr{\langle \gradient {e}_1, -\sum_{j=1}^{n-1} k_j {e}_j \gradient {e}_j \rangle_g \\ \vdots \\ \langle \gradient {e}_{n-1}, -\sum_{j=1}^{n-1} k_j {e}_j \gradient {e}_j \rangle_g},
		\end{split}
	\end{equation}
	where the arguments $\xi(t)$ are omitted for simplicity.
	
	We adopt the common convention in differential manifold theory that a vector field with a subscript represents the vector at the point represented by the subscript. Let $\tilde{\vf}$ denote the vector field defined on $\set{U}\times \mathbb{S}^1$ such that 
	\begin{equation} \label{eq_tildevf}
		\tilde{\vf}_{\Gamma({\xi})} = \difrntl \Gamma(\vf_{{\xi}})		
	\end{equation}
	for any point ${\xi} \in \set{V}$, where $\difrntl \Gamma$ is the tangent map of $\Gamma$. Since $\Gamma$ is a $C^2$-diffeomorphism, one can obtain information about $\vf$ by studying $\tilde{\vf}$. %
	One can calculate that
	\begin{multline} \label{eq13}
		\left( \sum_{i=1}^{n-1} k_i \pi_i^2(\cdot) \right) \circ \Gamma(\cdot)  %
		= \sum_{i=1}^{n-1} k_i (\pi_i^2 \circ \Gamma) (\cdot) 	= \sum_{i=1}^{n-1} k_i e_i^2 (\cdot) .
	\end{multline}
	By regarding a tangent vector as the velocity of a curve \cite[pp. 68-70]{lee2015introduction}, we have		
		\begin{equation} \label{eq_tangentvelocity}
			\vf\left( \sum_{i=1}^{n-1} k_i e_i^2(\cdot) \right) = \dt (\transpose{e(\xi(t))} K e(\xi(t))) |_{t=0},
		\end{equation}
	where $\xi: (-\epsilon, \epsilon) \to \set{V}$ is the trajectory of \eqref{eq1} with $\xi(0) \in \inv{\Gamma}(\partial \overline{\set{E}} \times \mathbb{S}^1)$ and $\dot{\xi}(0)=\vf_{\xi(0)}$. Let $V(\xi) \defeq \transpose{e(\xi)} K e(\xi)$, then from \eqref{eq_tangentvelocity}, we have
	\begin{multline} \label{eq18}
			 \dt V(\xi) = \dt (\transpose{e(\xi(t))} K e(\xi(t))) |_{t=0} \\ 
			\stackrel{\eqref{eq_e_dot}}{=}  \; 2 \transpose{ \matr{\langle \gradient {e}_1, -\sum_{j=1}^{n-1} k_j {e}_j \gradient {e}_j \rangle_g \\ \vdots \\ \langle \gradient {e}_{n-1}, -\sum_{j=1}^{n-1} k_j {e}_j \gradient {e}_j \rangle_g}} \matr{k_1 {e}_1 \\ \vdots \\ k_{n-1} {e}_{n-1}} \\
			= - 2 \left\langle \sum_{j=1}^{n-1} k_j {e}_j \gradient {e}_j, \sum_{j=1}^{n-1} k_j {e}_j \gradient {e}_j \right\rangle_g \le 0.
	\end{multline}
	Therefore, we have	
	$
		\tilde{\vf}\left( \sum_{i=1}^{n-1} k_i \pi_i^2(\cdot) \right) = \vf \left( \left( \sum_{i=1}^{n-1} k_i \pi_i^2(\cdot) \right) \circ \Gamma \right)  \overset{\eqref{eq13}}{=} \vf\left( \sum_{i=1}^{n-1} k_i e_i^2(\cdot) \right)  \overset{\eqref{eq_tangentvelocity},\eqref{eq18}}{=} - 2 \norm{ l(\cdot) }^2 \le 0,
	$
	where $l(\cdot) \defeq \sum_{j=1}^{n-1} k_j {e}_j(\cdot) \gradient {e}_j(\cdot)$. Note that for any point $\xi \in \manifold$, if $\norm{ l(\xi) }=0$, then either $\gradient {e}_j(\xi), j=1,\dots,n-1$, are linearly dependent, or they are linearly independent, but all  ${e}_j(\xi)$ equal zero. In the former case, the first term of \eqref{new_gvf} satisfies $\orthoterm(\xi)=0$, and thus $\vf(\xi)=0$, leading to $\xi \in \set{C}$. In the latter case, it is obvious that $\xi \in \set{P}$. Therefore, we have
	\begin{equation} \label{eq:pointinward}
		\tilde{\vf}_{\xi}\left( \sum_{i=1}^{n-1} k_i \pi_i^2(\cdot) \right) < 0, \quad \forall \xi \in \partial \overline{\set{E}} \times \mathbb{S}^1
	\end{equation} 
	as there are all regular points on $\inv{\Gamma}(\partial \overline{\set{E}} \times \mathbb{S}^1) \subseteq \set{V} \setminus \set{P}$, and 
	\begin{equation} \label{eq:invariant set}
		\tilde{\vf}_{\xi}\left( \sum_{i=1}^{n-1} k_i \pi_i^2(\cdot) \right) \le 0,  \quad \forall \xi \in \overline{\set{E}} \times \mathbb{S}^1
	\end{equation} 
	where the equality is taken only for points $\xi \in \Gamma(\set{P}) \approx \{ 0 \} \times \mathbb{S}^1$. 
	
	Given an initial condition $\zeta(0)\in\set{S}\subseteq \set{V}$, let $\zeta: \mbr[]_{\ge 0} \to \manifold$ be the solution to \eqref{eq1}. Since $\set{V}$ is an open subset in $\manifold$, there is some $\epsilon>0$ such that $\zeta(t) \in \set{V}$ for $t \in (-\epsilon, \epsilon)$.	Let 
	$
	\tilde{\zeta} \defeq \Gamma \circ \zeta |_{(-\epsilon,\epsilon)}
	$
	and therefore, by \eqref{eq_tildevf}, we have
	$
		\dt \tilde{\zeta} =\tilde{\vf} \circ \tilde{\zeta}\label{eq:X'system}
	$
	on $\set{U}\times \mathbb{S}^1$ with $\tilde{\zeta}(0)\in\partial\overline{\set{E}}\times \mathbb{S}^1$.	According to \eqref{eq:pointinward}, one has 
	$
	\dt \left( \sum_{i=1}^{n-1} k_i \pi_i^2(\cdot) \right) \circ\tilde{\zeta}(t)|_{t=0}=\tilde{\vf}\left( \sum_{i=1}^{n-1} k_i \pi_i^2(\cdot) \right)<0,
	$
	and hence by \eqref{eq:invariant set}, we have $\left( \sum_{i=1}^{n-1} k_i \pi_i^2(\cdot) \right) \circ \tilde{\zeta}(t)<R$ for any $t\in(0,\epsilon)$, where recall that $R$ is defined in \eqref{note_eqD}. Thus, $\tilde{\zeta}\big( (0,\epsilon) \big)\subseteq \set{E}\times \mathbb{S}^1$. Therefore, $\zeta \big( [0,\epsilon) \big)\subseteq e_{\set{V}}^{-1}(\set{E})$ and this shows that $\set{S}$ is the exit set of $\set{W}$. 
	
	\textbf{Proof of Claim \ref{claim3.2}:} Suppose $\zeta(0) \in e_{\set{V}}^{-1}(\overline{\set{E}})$. By \eqref{eq:invariant set} and using the same reasoning as above, we have $\left( \sum_{i=1}^{n-1} k_i \pi_i^2(\cdot) \right)\circ\tilde{\zeta}(t)\leq R$ for any $t\in[0,\epsilon)$, and thus $\tilde{\zeta}[0,\epsilon)\subseteq\overline{\set{E}}\times \mathbb{S}^1$. By the compactness of $\overline{\set{E}}\times \mathbb{S}^1$, one can show that there is an extension of the solution $\tilde{\zeta}$ of \eqref{eq:X'system} on $(-\epsilon,+\infty)$ such that $\tilde{\zeta}(0,\infty)\subseteq \set{E}\times \mathbb{S}^1$  \cite[Theorem 3.3]{khalil2002nonlinear}.	Then due to the uniqueness of the solution to the Cauchy problem of \eqref{eq1}, $\zeta=\Gamma^{-1}\circ\tilde{\zeta}$ is defined on $(-\epsilon,+\infty)$ and thus $\zeta(0,\infty) \subseteq \Gamma^{-1}(\set{E}\times \mathbb{S}^1)=e_{\set{V}}^{-1}(\set{E})$. Therefore, $e_{\set{V}}^{-1}(\overline{\set{E}})$ is an invariant set of the flow $\Psi$. The same argument can be applied to the case where $\zeta(0) \in e_{\set{V}}^{-1}(\set{E})$ and show that $e_{\set{V}}^{-1}(\set{E})$ is also an invariant set.
	
	\textbf{Proof of Claim \ref{claim3.3}:} By the previous argument, $e_{\set{V}}^{-1}(\overline{\set{E}})$ is compact and positively invariant. Also note that $\dot{V}(\xi) \le 0$ in $e_{\set{V}}^{-1}(\overline{\set{E}})$ from \eqref{eq18}. Therefore, according to the LaSalle's invariance principle \cite[Theorem 4.4]{khalil2002nonlinear}, the trajectory $\xi(t)$ starting from $e_{\set{V}}^{-1}(\overline{\set{E}})$ converges asymptotically to the largest invariant set in $\{\xi \in e_{\set{V}}^{-1}(\overline{\set{E}}) : \dot{V}(\xi)=0\}=\set{P}$ as $t \to \infty$, and it is obvious that the largest invariant set is the desired path $\set{P}$.   
	
	\textbf{Proof of Claim \ref{claim3.4}:} This is justified by the previous argument that $\zeta(0,\infty) \subseteq \Gamma^{-1}(\set{E}\times \mathbb{S}^1)=e_{\set{V}}^{-1}(\set{E})$.
\end{proof}

\subsection{Proof of Lemma \ref{note_lemma4}} \label{app3}
\begin{proof}
	\textbf{Step 1 (Prove $\vf$ transverse to $\set{S}$): } We first show that ${\rm span}\{\vf_{{\xi}}\}\oplus T_{{\xi}} \set{S} = T_{{\xi}} \manifold$ for ${\xi} \in \set{S}$, where $\oplus$ is the direct sum; that is, the vector field $\vf$ is transverse to $\set{S}$ in $\manifold$. Since $\set{S}$ is homeomorphic to $\partial\overline{\set{E}}\times \mathbb{S}^1$ by Corollary \ref{coroll1}, $\set{S}$ is an $(n-1)$-dimensional submanifold in $\manifold$. For any ${\xi} \in \set{S}$, we have $ \difrntl e_{{\xi}} (T_{\xi} \set{S})=T_{e({\xi})} \partial \overline{\set{E}}$, where $\difrntl e_{{\xi}}$ is the tangent map of $e$ at ${\xi}$. Then \eqref{eq:pointinward} implies that $\vf_{{\xi}} \notin T_{{\xi}} \set{S}$ for ${\xi} \in \set{S}$, and hence ${\rm span}\{\vf_{{\xi}}\} \oplus T_{{\xi}} \set{S} = T_{{\xi}}\manifold$ for ${\xi} \in \set{S}$; that is, the vector field $\vf$ is transverse to $\set{S}$ in $\manifold$.	
	
	\textbf{Step 2 (Prove $\Psi$ is open): } Since ${\rm span}\{\vf_{{\xi}}\} \oplus T_{{\xi}} \set{S}=T_{{\xi}}\manifold$ for any point ${\xi} \in \set{S}$ as shown above, the tangent map $\difrntl \Psi_{(0,{\xi})}$ of the flow $\Psi(\cdot,\cdot)$ is a surjection from $T_{(0,{\xi})}(\mbr[] \times \set{S} )$ onto $T_{\Psi(0,{\xi})}\manifold$ for each point $(0,{\xi})\in \mbr[] \times \set{S}$. Therefore, $\Psi(\cdot,\cdot)$ is locally an open map from some open neighbourhood $\set{O}$ of $(0,{\xi})$. Thus, $\Psi(\set{O}) \defeq \{\Psi^{t}(q) : (t,q)\in \set{O}\}$ is open in $\manifold$. For any fixed $t_{0}\in\mbr[]$, denote by $t_{0}+\set{O}$ the set	$\{(t+t_{0},q) : (t,q)\in \set{O}\}$, which is an open neighbourhood of $(t_{0},{\xi})$. Therefore, $\Psi(t_{0}+\set{O}) = \{\Psi^{t_{0}+t}(q)=\Psi^{t_{0}} (\Psi^{t}(q)) : (t,q)\in \set{O}\} =\Psi^{t_{0}}\left( \{\Psi^{t}(q) : (t,q)\in \set{O}\} \right) = \Psi^{t_{0}}\left( \Psi(\set{O}) \right)$
	is open since $\Psi^{t_{0}}(\cdot)$ is a $C^2$-diffeomorphism of $\manifold$. Thus, $\Psi(\cdot,\cdot)$ is an open map from $\mbr[]\times \set{S}$ to $\manifold$.
	
	\textbf{Step 3 (Prove $\Psi$ is injective): } Suppose $\Psi^{t}(p)=\Psi^{t'}(q)$ for	some $p, q \in \set{S}$ and $t,t' \in \mbr[]$. If $t=t'$, then $p=q$ due to the uniqueness of the solution to the Cauchy problem in \eqref{eq1}. Now assume that $t<t'$, then $p=\Psi^{t'-t}(q)$. According to Lemma \ref{note_lemma3} \ref{claim3.4}, $p \in e_{\set{V}}^{-1}(\set{E})$, which contradicts the condition that $p \in \set{S}$. A similar contradiction arises if $t>t'$. Therefore, we have $t=t'$ and $p=q$, justifying that $\Psi(\cdot,\cdot)$ is an injection from $\mbr[]\times \set{S}$ to $\manifold$.
	
	\textbf{Step 4 (Prove $\Psi$ is a homeomorphism): } We have proved above that the flow $\Psi : \mbr[] \times \set{S}  \to \manifold$ of \eqref{eq1} is an open injection. Now if we restrict the codomain of $\Psi$ to its image $\Psi^{(-\infty,+\infty)}(\set{S})$, then the map $\Psi : (-\infty,+\infty) \times \set{S} \to \Psi^{(-\infty,+\infty)}(\set{S})$ is continuous, bijective, and open, hence a homeomorphism \cite[Theorem A.38 (c)]{lee2015introduction}.
\end{proof}

\subsection{Proof of Lemma \ref{note_lemma5} } \label{app4}
\begin{proof}
	\textbf{Proof of Claim \ref{claim5.1}:} We first show that $\set{W}^{\circ}=\Psi^{(-\infty,0]}(\set{S})$. According to Definition \ref{note_def3} and \eqref{eqwcirc}, for any ${\xi} \in \set{W}^{\circ}$, there exists some $\tau>0$ such that $\Psi^{\tau}({\xi}) \in e_{\set{V}}^{-1}(\set{E})$. Let 
	$
	a \defeq \inf\{t\in(0,\tau) : \Psi^{[t,\tau]}({\xi}) \subseteq e_{\set{V}}^{-1}(\set{E})\} \ge 0,
	$
	and then $\Psi^{a}({\xi}) \in e_{\set{V}}^{-1}(\partial \overline{\set{E}})$, and therefore ${\xi} \in \Psi^{-a}(e_{\set{V}}^{-1}(\partial \overline{\set{E}}))=\Psi^{-a}(\set{S}) \subseteq \Psi^{(-\infty,0]}(\set{S})$. Thus, $\set{W}^{\circ} \subseteq \Psi^{(-\infty,0]}(\set{S})$. Conversely, suppose ${\xi} \in \Psi^{(-\infty,0]}(\set{S})$. Then there exists some $t \le 0$ and $y \in \set{S}$ such that $\Psi^{t}(y)={\xi}$, or $y=\Psi^{-t}({\xi})$. Since $\set{S}$ is the exit set of $\set{W}$ by Lemma \ref{note_lemma3} \ref{claim3.1}, there exists a positive constant $\delta$ such that $\Psi^{(0,\delta)}(y) \subseteq e_{\set{V}}^{-1}(\set{E})$. Therefore, we have $\Psi^{-t + \delta/2}({\xi}) \in e_{\set{V}}^{-1}(\set{E})$. In view of the definition of $\set{W}^{\circ}$ in \eqref{eqwcirc}, it follows that $\xi \in \set{W}^{\circ}$, hence $\Psi^{(-\infty,0]}(\set{S}) \subseteq \set{W}^{\circ}$. To sum up, we have $\set{W}^{\circ}=\Psi^{(-\infty,0]}(\set{S})$.
	
	Now we show that $\set{W}^{\circ}$ is open in $\set{W}$ by showing that $\set{W}^{\circ}=\Psi^{(-\infty,+\infty)}(\set{S})\cap \set{W}$. Since $e_{\set{V}}^{-1}(\set{E})$ is an invariant set of the flow by Lemma \ref{note_lemma3} \ref{claim3.2}, $\Psi^{(0,+\infty)}(q) \subseteq e_{\set{V}}^{-1}(\set{E})$ for every $q \in \set{S}$, and therefore, $\Psi^{(0,+\infty)}(\set{S}) \cap \set{W}=\emptyset$. So $\Psi^{(-\infty,+\infty)}(\set{S}) \cap \set{W} = \Psi^{(-\infty,0]}(\set{S})\cap \set{W} = \set{W}^{\circ} \cap \set{W}=\set{W}^{\circ}$. Since $\Psi^{(-\infty,+\infty)}(\set{S})$ is open in $\manifold$ by Lemma \ref{note_lemma4}, $\set{W}^{\circ}$ is open in $\set{W}$. 
	
	\textbf{Proof of Claim \ref{claim5.2}:} By Lemma \ref{note_lemma4}, $\Psi : (-\infty,+\infty) \times \set{S} \to \Psi^{(-\infty,+\infty)}(\set{S})$	is a homeomorphism. Hence $\Psi^{(\cdot)}(\cdot)$ is a homeomorphism from $(-\infty,0] \times \set{S}$ to $\set{W}^{\circ}$.
\end{proof}
\begin{figure}[tb]
	\centering
	\begin{tikzpicture}
		\matrix (m) [matrix of math nodes,row sep=1em,column sep=6em,minimum width=2em]
		{ \overline{\set{E}}\times \mathbb{S}^1 & e_{\set{V}}^{-1}(\overline{\set{E}}) & \set{W}^{\circ}\\ 
			(-\infty, 0] \times \partial \overline{\set{E}}\times \mathbb{S}^1 & {} & \set{W}^{\circ} \\ };
		\path[-stealth]
		(m-1-1) edge node [above] {$\Gamma^{-1}$} (m-1-2)
		(m-1-2) edge node [above] {$\Psi^{t}$} (m-1-3)
		(m-2-1) edge node [above] {$\Upsilon(\cdot,\cdot,\cdot)=\Psi^{(\cdot)} \circ \Gamma^{-1}|_{\partial\overline{\set{E}} \times \mathbb{S}^1}(\cdot,\cdot)$} (m-2-3);
	\end{tikzpicture}
	\caption{The map $\Upsilon(t,\xi,\zeta)=\Psi^{t} \circ \Gamma^{-1}|_{\partial\overline{\set{E}} \times \mathbb{S}^1}(\xi,\zeta)$.  }
	\label{fig: GammaLambda}
\end{figure}
\begin{figure}[tb]
	\centering
	\begin{tikzpicture}
		\matrix (m) [matrix of math nodes,row sep=1em,column sep=4em,minimum width=2em]
		{
			\set{X}  & {} \\
			\set{X} / \sim & \set{W}^{\circ} \cup e_{\set{V}}^{-1}(\overline{\set{E}}) \\};
		\path[-stealth]
		(m-1-1) edge node [left] {$\mathrm{pr}$} (m-2-1) 
		(m-2-1) edge node [below] {$\Lambda$} (m-2-2)
		(m-1-1) edge node [above right] {$\Upsilon\sqcup\Gamma^{-1}$} (m-2-2);
	\end{tikzpicture}
	\caption{Since $\mathrm{pr}$ is a quotient map and $\Upsilon \sqcup \Gamma^{-1}$ is continuous, $\Lambda$ is continuous. }
	\label{fig:quotientcont1}
	\vspace{-1em}
\end{figure}
\begin{figure}[tb]
	\centering
	\begin{tikzpicture}
		\matrix (m) [matrix of math nodes,row sep=1em,column sep=4em,minimum width=2em]
		{
			(-\infty,0]\times\partial\overline{\set{E}} &  \mbr[n-1] \setminus \set{E} \\
			\set{X}  & {} \\
			\set{X} / \sim & \mbr[n-1]\times \mathbb{S}^1 \\};
		\path[-stealth]
		(m-1-1) edge node [above] {$\Xi$} (m-1-2) 
		(m-2-1) edge node [left] {$\mathrm{pr}$} (m-3-1) 
		(m-3-1) edge node [below] {$\tilde{\Lambda}$} (m-3-2)
		(m-2-1) edge node [above right] {$\Xi \times \identity_{\mathbb{S}^1}\sqcup \identity_{\overline{\set{E}}}\times \identity_{\mathbb{S}^1}$} (m-3-2);
	\end{tikzpicture}
	\caption{Since $\mathrm{pr}$ is a quotient map and $\Xi \times \identity_{\mathbb{S}^1}\sqcup \identity_{\overline{\set{E}}}\times \identity_{\mathbb{S}^1}$ is continuous, $\tilde{\Lambda}$ is continuous. }
	\label{fig:quotientcont2}
\end{figure}

\subsection{Proof of Proposition \ref{note_prop3}} \label{app5}
\begin{proof}
	By Lemma \ref{note_lemma5} \ref{claim5.2}, $(t, q) \mapsto \Psi^{t}(q)$ is a homeomorphism from $(-\infty,0] \times \set{S}$	to $\set{W}^{\circ}$. In addition, $\set{S}$ is homeomorphic to $\partial\overline{\set{E}}\times \mathbb{S}^1$ with the homeomorphism $\Gamma^{-1}|_{\partial\overline{\set{E}} \times \mathbb{S}^1}$ by Corollary \ref{coroll1}. Therefore,
	\begin{equation*}
		\begin{split}
			\Upsilon: (-\infty,0]\times\partial\overline{\set{E}}\times \mathbb{S}^1 &\to \set{W}^{\circ}  \\
			(t,\xi,\zeta) &\mapsto \Psi^{t} \circ \Gamma^{-1}|_{\partial\overline{\set{E}} \times \mathbb{S}^1}(\xi,\zeta)
		\end{split}
	\end{equation*} 
	is a homeomorphism, where $\Gamma^{-1}: \overline{\set{E}}\times \mathbb{S}^1 \to e_{\set{V}}^{-1}(\overline{\set{E}})$ is also a homeomorphism (see Fig. \ref{fig: GammaLambda}). Let
	$
	\set{X} \defeq  (-\infty,0] \times \partial \overline{\set{E}} \times \mathbb{S}^1 \sqcup \overline{\set{E}} \times \mathbb{S}^1
	$
	with the disjoint union topology \cite[p. 64]{lee2010topologicalmanifolds}. We define a new topological space 
	$
	\set{X} / \sim, 
	$
	where $\sim$ is the equivalence relation that identifies $(0,\xi,\zeta)$ in $\{0\} \times \partial\overline{\set{E}}\times \mathbb{S}^1$ with $(\xi,\zeta)$ in $\partial\overline{\set{E}}\times \mathbb{S}^1$. The natural projection
	$
	\mathrm{pr} :  \set{X}  \to \set{X}/ \sim
	$
	is a quotient map \cite[Chapter 2, Section 22]{munkres2013topology}. Note that $e_{\set{V}}^{-1}(\overline{\set{E}})$ and $\set{W}^{\circ}$ are subspaces of $e_{\set{V}}^{-1}(\overline{\set{E}})\cup \set{W}^{\circ}$. Therefore, the map
	\begin{align*}
		\Upsilon\sqcup\Gamma^{-1}: \set{X}  &\to \set{W}^{\circ} \cup e_{\set{V}}^{-1}(\overline{\set{E}}) \\
		(-\infty,0] \times \partial \overline{\set{E}} \times \mathbb{S}^1\ni(t,\xi,\zeta) &\mapsto \Upsilon(t,\xi,\zeta)\\
		\overline{\set{E}}\times \mathbb{S}^1\ni(\xi,\zeta) &\mapsto  \Gamma^{-1}(\xi,\zeta)
	\end{align*}
	is continuous. By construction, there exists the unique map
	$
	\Lambda : \set{X}/ \sim \,\to\,  \set{W}^{\circ} \cup e_{\set{V}}^{-1}(\overline{\set{E}})
	$ 
	such that $\Lambda \circ \mathrm{pr} = \Upsilon\sqcup\Gamma^{-1}$ (see Fig. \ref{fig:quotientcont1}). Due to the property of a quotient map \cite[Theorem 22.2]{munkres2013topology}, the map $\Lambda$ is continuous. One can check that $\Lambda$ is a bijection. 
	In addition, $\Lambda$ is an open map. This is justified as follows. First, both $(-\infty,0]\times\partial\overline{\set{E}}\times \mathbb{S}^1$ and $\overline{\set{E}}\times \mathbb{S}^1$ are topological manifolds with boundaries $\{0\}\times\partial\overline{\set{E}}\times \mathbb{S}^1$ and $\partial\overline{\set{E}}\times \mathbb{S}^1$ respectively. Note that the boundaries are homeomorphic to each other in a natural way which is given by the equivalent relation $\sim$; i.e., $(0,\xi,\zeta) \mapsto (\xi,\zeta)$ is a natural homeomorphism. Then it follows from applying the technique of attaching manifolds together along their boundaries \cite[Theorem 3.79]{lee2010topologicalmanifolds} that $\set{X}/ \sim$ is an $n$-manifold. Second, $ \set{W}^{\circ} \cup e_{\set{V}}^{-1}(\overline{\set{E}})=\Psi^{(-\infty,+\infty)}(\set{S}) \cup e_{\set{V}}^{-1}(\set{E})$, which is an open subset in $\manifold$, and hence it is also an $n$-manifold. Thus, $\Lambda$ is a continuous injection between two (boundaryless) $n$-manifolds, and hence it is open \cite[Theorem 36.5]{munkres2018elements}. Therefore, $\Lambda$ is a homeomorphism, and it remains to show that $\set{X}/ \sim$ is homeomorphic to $\mbr[n-1]\times \mathbb{S}^1$. 
	
	Note that $\mbr[n-1]=\overline{\set{E}}\cup(\mbr[n-1] \setminus \set{E})$. Define	$\Xi:(-\infty,0]\times\partial\overline{\set{E}} \to \mbr[n-1] \setminus \set{E}$ by $(t,\xi) \mapsto \xi - t \xi.	$
	It is obvious that $\Xi$ is a homeomorphism. Applying the same argument above, one can show that
	$
	\Xi \times \identity_{\mathbb{S}^1}\sqcup \identity_{\overline{\set{E}}}\times \identity_{\mathbb{S}^1}: \set{X} \to \mbr[n-1]\times \mathbb{S}^1,
	$
	induces a homeomorphism $\tilde{\Lambda}$ between $\set{X}/ \sim$ and $\mbr[n-1]\times \mathbb{S}^1$ (see Fig. \ref{fig:quotientcont2}).
\end{proof}

\bibliographystyle{IEEEtran}
\bibliography{ref}

\end{document}